%% file: main.tex
\documentclass[lettersize,journal]{IEEEtran}
\usepackage{amsmath,amsfonts}
\usepackage{mathrsfs}
\usepackage{algorithmic}
\usepackage{algorithm}
\usepackage{array}
\usepackage[caption=false,font=normalsize,labelfont=sf,textfont=sf]{subfig}
\usepackage{textcomp}
\usepackage{stfloats}
\usepackage{url}
\usepackage{verbatim}
\usepackage{color,soul}
\usepackage{graphicx}
\usepackage{hyperref}
\usepackage{cite}
\usepackage[nolist]{acronym}
\usepackage{mathtools}
\usepackage{amssymb}
\usepackage{stfloats}
\usepackage{makecell}
\usepackage{nccmath}
\acrodefplural{HPA}[HPAs]{high power amplifiers}
\usepackage{yhmath}

\input{acronyms.tex}

\DeclareMathOperator{\ISMR}{ISMR}
\DeclareMathOperator{\trace}{tr}

\DeclareMathOperator{\SR}{SR}

\DeclareMathOperator{\dB}{dB}

\DeclareMathOperator{\rank}{rank}
\DeclareMathOperator{\EVD}{EVD}
\DeclareMathOperator{\dBm}{dBm}
\DeclareMathOperator{\dBi}{dBi}
\DeclareMathOperator{\meters}{m}
\DeclareMathOperator{\GHz}{GHz}
\DeclareMathOperator{\dBmpHz}{dBm/Hz}
\DeclareMathOperator{\bpspHz}{bits/sec/Hz}

\newcommand*\HSI{\pmb{H}_{\mathrm{SI}}}

\begin{document}
\title{Sum Secrecy Rate Maximization for Full Duplex ISAC Systems}
\author{Aleksandar Boljevi\'c, Ahmad Bazzi and Marwa Chafii 
\thanks{
Aleksandar Boljevi\'c is with the NYU Tandon School of Engineering, Brooklyn, 11201, NY, US and the Engineering Division, New York University (NYU) Abu Dhabi, 129188, UAE
(email: \href{ab10863@nyu.edu}{ab10863@nyu.edu}).

Ahmad Bazzi and Marwa Chafii are with the Engineering Division, New York University (NYU) Abu Dhabi, 129188, UAE and NYU WIRELESS,
NYU Tandon School of Engineering, Brooklyn, 11201, NY, USA
(email: \href{ahmad.bazzi@nyu.edu}{ahmad.bazzi@nyu.edu}, \href{marwa.chafii@nyu.edu}{marwa.chafii@nyu.edu}).
.}
\thanks{Manuscript received xxx}}
\markboth{submitted, 2024}%
{Shell \MakeLowercase{\textit{et al.}}: A Sample Article Using IEEEtran.cls for IEEE Journals}


\maketitle

\begin{abstract}
\input{sections/abstract.tex}
\end{abstract}

\section{Introduction}
\label{sec:introduction}
\input{sections/introduction.tex}

\section{System Model}
\label{sec:system-model}
\input{sections/system-model.tex}

\vspace{-0.2cm}
\section{Secure FD-ISAC Sum Secrecy Rate Optimization}
\label{sec:optimization-framework}
\input{sections/optimization-framework.tex}

\vspace{-0.1cm}
\section{Iterative Joint Taylor-BCCD type method}
\label{sec:algorithmic-derivation}
\input{sections/algorithmicderivation}
\input{sections/algorithm.tex}

\section{Simulation Results}
\label{sec:simulation-results}
\input{sections/simulation-results.tex}

\section{Conclusion}
\label{sec:conclusion}
\input{sections/conclusion.tex}

\vspace{-0.2cm}
\appendices
\section{Proof of Lemma 1}
\label{sec:appendix1}
\input{sections/appendix1.tex}



\section{Acknowledgment}
This work has been supported by Tamkeen and the Center for Cybersecurity under the NYU Abu  Dhabi Research Institute Award G1104.

\bibliographystyle{IEEEtran}
\bibliography{refs}

\vfill

\end{document}

%% file: acronyms.tex
\begin{acronym}
	\acro{mmWave}{milli-meter wave}
    \acro{MIMO}{multiple-input, multiple-output}
    \acro{AWGN}{additive white Gaussian noise}
    \acro{KPI}{key performance indicator}
    \acro{KPIs}{key performance indicators}  
    \acro{D2D}{device-to-device}
    \acro{ISAC}{integrated sensing and communication}
    \acro{DFRC}{dual-functional radar and communication}
    \acro{AoA}{angle of arrival}
    \acro{AoAs}{angles of arrival}
    \acro{ToA}{time of arrival}
    \acro{EVM}{error vector magnitude}
    \acro{CPI}{coherent processing interval}
    \acro{eMBB}{enhanced mobile broadband}
    \acro{URLLC}{ultra-reliable low latency communications}
    \acro{mMTC}{massive machine type communications}
	\acro{QCQP}{quadratic constrained quadratic programming}
	\acro{BnB}{branch and bound}
	\acro{SER}{symbol error rate}
	\acro{LFM}{linear frequency modulation}
	\acro{BS}{base station}
	\acro{UE}{user equipment}
	\acro{DL}{downlink}
	\acro{UL}{uplink}
	\acro{CS}{compressed sensing}
	\acro{PR}{passive radar}
	\acro{LMS}{least mean squares}
	\acro{NLMS}{normalized least mean squares}
	\acro{RIS}{reconfigurable intelligent surface}
    \acro{UAV}{unmanned aerial vehicle}
    \acro{STARS}{simultaneously transmitting and reflecting surface}
	\acro{RISs}{reconfigurable intelligent surfaces}
	\acro{IRS}{intelligent reflecting surface}
	\acro{LISA}{large intelligent surface/antennas}
	\acro{OFDM}{orthogonal frequency-division multiplexing}
	\acro{EM}{electromagnetic}
	\acro{MSE}{mean squared error}
	\acro{XR}{extended reality}
	\acro{SNR}{signal-to-noise ratio}
	\acro{SRP}{successful recovery probability}
	\acro{CDF}{cumulative distribution function}
	\acro{ULA}{uniform linear array}
	\acro{RSS}{received signal strength}
	\acro{SNDR}{signal-to-noise-and-distortion ratio}
	\acro{DPI}{direct path interference}
	\acro{RPI}{reflected path interference}
	\acro{DE}{direct echo}
	\acro{RE}{RIS-resulted echo}
	\acro{PI}{path interference}
	\acro{ADC}{analog-to-digital converter}
	\acro{RF}{radio frequency}
	\acro{BCD}{block coordinate descent}
	\acro{BCCD}{block cyclic coordinate descent}
	\acro{RCG}{Riemannian conjugate gradient}
	\acro{SDP}{semi-definite program}
	\acro{EVD}{eigenvalue decomposition}
	\acro{RCS}{radar cross-section}
	\acro{STAR}{simultaneous transmitting and reflecting}
	\acro{LNA}{low noise amplifier}
    \acro{NOMA}{non-orthogonal multiple access}
    \acro{FM}{frequency modulation}
    \acro{SINR}{signal to interference plus noise ratio}
    \acro{V2X}{Vehicle-to-Everything}
    \acro{NLoS}{non-line-of-sight}
    \acro{LoS}{line-of-sight}
    \acro{AN}{artificial noise}
	\acro{SI}{self interference}
    \acro{MIMO}{multiple-input, multiple-output}
    \acro{AWGN}{additive white Gaussian noise}
    \acro{KPI}{key performance indicator}
    \acro{KPIs}{key performance indicators}  
    \acro{D2D}{device-to-device}
    \acro{ISAC}{integrated sensing and communications}
    \acro{DFRC}{dual-functional radar and communication}
    \acro{AoA}{angle of arrival}
    \acro{AoAs}{angles of arrival}
    \acro{MU-MIMO}{multi-user, multiple-input, multiple-output}
    \acro{ToA}{time of arrival}
    \acro{PCA}{principle component analysis}
    \acro{EVM}{error vector magnitude}
    \acro{CPI}{coherent processing interval}
    \acro{eMBB}{enhanced mobile broadband}
    \acro{URLLC}{ultra-reliable low latency communications}
    \acro{mMTC}{massive machine type communications}
	\acro{QCQP}{quadratic constrained quadratic programming}
	\acro{BnB}{branch and bound}
	\acro{SER}{symbol error rate}
	\acro{LFM}{linear frequency modulation}
	\acro{BS}{base station}
	\acro{SLNR}{signal-to-leakage-plus-noise ratio}
	\acro{UE}{user equipment}
	\acro{DL}{deep learning}
	\acro{CS}{compressed sensing}
	\acro{PR}{passive radar}
	\acro{LMS}{least mean squares}
	\acro{NLMS}{normalized least mean squares}
	\acro{ZF}{zero forcing}
	\acro{RIS}{reconfigurable intelligent surface}
	\acro{RISs}{reconfigurable intelligent surfaces}
	\acro{IRS}{intelligent reflecting surface}
	\acro{IRSs}{intelligent reflecting surfaces}
	\acro{NOMA}{non-orthogonal multiple access}
	\acro{TWRN}{two-way relay networks}
	\acro{LISA}{large intelligent surface/antennas}
	\acro{LISs}{large intelligent surfaces}
	\acro{OFDM}{orthogonal frequency-division multiplexing}
	\acro{EM}{electromagnetic}
	\acro{ISMR}{integrated sidelobe to mainlobe ratio}
	\acro{MSE}{mean squared error}
	\acro{SNR}{signal-to-noise ratio}
	\acro{IoT}{internet of things}
	\acro{SRP}{successful recovery probability}
	\acro{CDF}{cumulative distribution function}
	\acro{ULA}{uniform linear array}
	\acro{RSS}{received signal strength}
	\acro{SU}{single user}
	\acro{MU}{multi-user}
	\acro{CSI}{channel state information}
	\acro{OFDM}{orthogonal frequency-division multiplexing}
	\acro{DL}{downlink}
	\acro{i.i.d}{independently and identically distributed}
	\acro{UL}{uplink}
	\acro{DFT}{discrete Fourier transform}
    \acro{HPA}{high power amplifier}
    \acro{IBO}{input power back-off}
    \acro{MIMO}{multiple-input, multiple-output}
    \acro{PAPR}{peak-to-average power ratio}
    \acro{AWGN}{additive white Gaussian noise}
    \acro{KPI}{key performance indicator}
    \acro{FD}{full duplex}
    \acro{KPIs}{key performance indicators}  
    \acro{D2D}{device-to-device}
    \acro{ISAC}{integrated sensing and communications}
    \acro{DFRC}{dual-functional radar and communication}
    \acro{AoA}{angle of arrival}
    \acro{AoD}{angle of departure}
    \acro{ToA}{time of arrival}
    \acro{EVM}{error vector magnitude}
    \acro{CPI}{coherent processing interval}
    \acro{eMBB}{enhanced mobile broadband}
    \acro{URLLC}{ultra-reliable low latency communications}
    \acro{mMTC}{massive machine type communications}
	\acro{QCQP}{quadratic constrained quadratic programming}
	\acro{BnB}{branch and bound}
	\acro{SER}{symbol error rate}
	\acro{LFM}{linear frequency modulation}
	\acro{ADMM}{alternating direction method of multipliers}
	\acro{CCDF}{complementary cumulative distribution function}
	\acro{MISO}{multiple-input, single-output}
	\acro{CSI}{channel state information}
	\acro{LDPC}{low-density parity-check}
	\acro{BCC}{binary convolutional coding}
	\acro{IEEE}{Institute of Electrical and Electronics Engineers}
	\acro{ULA}{uniform linear antenna}
	\acro{B5G}{beyond 5G}
	\acro{MOOP}{multi-objective optimization problem}
	\acro{ML}{maximum likelihood}
	\acro{FML}{fused maximum likelihood}
	\acro{RF}{radio frequency}
	\acro{AM/AM}{amplitude modulation/amplitude modulation}
	\acro{AM/PM}{amplitude modulation/phase modulation}
	\acro{BS}{base station}
	\acro{MM}{majorization-minimization}
	\acro{LNCA}{$\ell$-norm cyclic algorithm}
	\acro{OCDM}{orthogonal chirp-division multiplexing}
	\acro{TR}{tone reservation}
	\acro{COCS}{consecutive ordered cyclic shifts}
	\acro{LS}{least squares}
	\acro{CVE}{coefficient of variation of envelopes}
	\acro{BSUM}{block successive upper-bound minimization}
	\acro{ICE}{iterative convex enhancement}
	\acro{SOA}{sequential optimization algorithm}
	\acro{BCD}{block coordinate descent}
	\acro{SINR}{signal to interference plus noise ratio}
	\acro{MICF}{modified iterative clipping and filtering}
	\acro{PSL}{peak side-lobe level}
	\acro{SDR}{semi-definite relaxation}
	\acro{KL}{Kullback-Leibler}
	\acro{ADSRP}{alternating direction sequential relaxation programming}
	\acro{MUSIC}{MUltiple SIgnal Classification}
	\acro{EVD}{eigenvalue decomposition}
	\acro{SVD}{singular value decomposition}
	\acro{LDGM}{low-density generator matrix}
	\acro{SAC}{sensing and communication}
    \acro{HD}{half duplex}
    \acro{DM}{directional modulation}
    \acro{CDA}{constellation decomposition array}
    \acro{IAB}{integrated access and back-haul}
    \acro{PLS}{physical layer security}
    \acro{THz}{terahertz}
    \acro{CU}{communication user}
    \acro{SID}{study item description}
    \acro{3GPP}{3rd Generation Partnership Project}
    \acro{CRB}{Cram\'er-Rao Bound}
    \acro{ISABC}{integrated sensing and backscatter communication}
    \acro{DoF}{degrees-of-freedom}
    \acro{IJTB}{iterative joint Taylor-BCCD}
\end{acronym}

%% file: sections/abstract.tex
In integrated sensing and communication (ISAC) systems, the target of interest may \textit{intentionally disguise itself as an eavesdropper}, enabling it to intercept and tap into the communication data embedded in the ISAC waveform.
The following paper considers a full duplex (FD)-ISAC system, which involves multiple malicious targets attempting to intercept both uplink (UL) and downlink (DL) communications between the dual-functional radar and communication (DFRC) base station (BS) and legitimate UL/DL communication users (CUs).
For this, we formulate an optimization framework that allows maximization of both UL and DL sum secrecy rates, under various power budget constraints for sensing and communications.
As the proposed optimization problem is non-convex, we develop a method called Iterative Joint Taylor-Block cyclic coordinate descent (IJTB) by proving essential lemmas that transform the original problem into a more manageable form. In essence, IJTB alternates between two sub-problems: one yields UL beamformers in closed-form, while the other approximates the solution for UL power allocation, artificial noise covariance, and DL beamforming vectors. This is achieved through a series of Taylor approximations that effectively \textit{"convexify"} the problem, enabling efficient optimization.
Simulation results demonstrate the effectiveness of the proposed solver when compared with benchmarking ones. Our findings reveal that the IJTB algorithm shows fast convergence, reaching stability within approximately $10$ iterations.
In addition, all benchmarks reveal a substantial decline in the sum secrecy rate, approaching zero as the eavesdropper distance reaches $17$ meters, underscoring their vulnerability in comparison to IJTB.
But even more, IJTB consistently adapts to varying eavesdropper distances, outperforming other methods, achieving a sum secrecy rate of $21.4 \bpspHz$ compared to $18.86 \bpspHz$ for the isotropic no-AN benchmark at a $50$ meter distance.
From a sensing point of view, IJTB continues to achieve high sum secrecy rates, particularly as the integrated sidelobe to mainlobe constraint becomes less stringent.

%% file: sections/introduction.tex
Wireless communication and radar sensing, two of the most successful applications of electromagnetic radiation, have historically been developed separately over the past century, each with distinct performance metrics and frequency bands \cite{richards2010principles}. However, with the continuous expansion of the communication frequency spectrum and the increasing desire to bridge the cyber and physical worlds through ubiquitous sensing, the landscape is shifting.
More precisely, beyond $5$G/$6$G, wireless systems are intended to provide a variety of high-accuracy sensing services, including indoor localization for robot navigation, Wi-Fi sensing for smart homes, machine learning for sensing \cite{10496165}, hybrid radar fusion (HRF) \cite{10417003}, and radar scanning for autonomous cars \cite{9705498}. In fact, \ac{SAC} systems are often built separately and operate in various frequency bands \cite{10494224}. Moreover, due to the widespread deployment of \ac{mmWave} technologies, communication signals in future wireless systems are expected to have high resolution in both the temporal and angle domains, allowing for high-accuracy sensing. As a result, simultaneous \ac{SAC} design is a favorable approach so that both systems share the same frequency band and hardware \cite{10416996}, therefore improving spectrum efficiency and lowering hardware costs. This encourages the research into \ac{ISAC}, as the aim is to efficiently utilize valuable radio resources and hardware for both \ac{SAC} purposes. Even more, time-division duplexing may be inefficient especially for demanding sensing and communication applications \cite{10463523}, thus, researchers have to look for more advanced, yet challenging integration schemes. \\
\indent Like any emerging field, \ac{ISAC} and $6$G in general are met with an extensive list of challenges, such as beyond a milli-second of latency \cite{10061453}. The authors in \cite{10041914} outline at least twelve of them, one of which is in part concerned with the role of security in \ac{ISAC}. 
Security has consistently been a primary concern for \ac{ISAC} systems \cite{10373185}. The sensing process necessitates full interaction of the signal with its environment, which increases the risk of eavesdropping. More specifically, what makes \ac{ISAC} especially vulnerable to security threats is that targets can use the \ac{ISAC} signal to decode communication information. Assuming the targets are malicious eavesdroppers, a dilemma arises: while the waveform design should direct radar power towards the eavesdropper’s direction, it should reveal the lowest possible communication rate in its direction. Simultaneously, it must achieve an acceptable communication rate towards legitimate \acp{CU}. Furthermore, \cite{9762838} outlines additional security challenges that are on the horizon, such as the tolerance for end-to-end latency, the large scale of massive machine type communications networks, extensive \ac{IoT} deployments, the extended lifespan of these deployments, and the diverse range of heterogeneous radio frequency technologies, among others.
It is worth noting that the \ac{3GPP} has initiated a \ac{SID} related to security aspects of \ac{ISAC} \cite{3gpp_sa3_2023}. In fact, security has already been identified as a standalone work task, deemed essential and of high priority for inclusion in the scope of release 19. But, inherent to \ac{ISAC}, it is crucial to study security and privacy across various functional and procedural aspects, including the authorization of service initiation, and \textit{the protection of the sensing signal and data collected}.
This work concentrates on the security concerns of \ac{ISAC}, with plans to address privacy concerns in future research.
\subsection{Existing work}
The work in \cite{9933849} utilizes a sequence of snapshots for robust and secure resource allocation in an ISAC system consisting of a multi-antenna DFRC BS serving multiple single-antenna legitimate users, while sensing potential single-antenna eavesdroppers. Similarly, the authors in \cite{10153696} consider a \ac{DL} secure ISAC system with a multi-antenna BS transmitting confidential messages to single-antenna CUs while simultaneously sensing targets.
Furthermore, \cite{10227884} explores the sensing-aided secure \ac{ISAC} system, where a dual-functioning \ac{BS} emits omnidirectional waveform in search of potential eavesdroppers and simultaneously sends confidential messages to \acp{CU}. The work intends to maximize the secrecy rate, while minimizing the eavesdroppers’ \ac{CRB}. 
Moreover, \cite{9857564} aims at designing an \ac{ISAC} beamformer that ensures secure communication rates, while generating a beampattern sufficiently close to the desired one, under the given transmit power budget constraint. 
Meanwhile, \cite{Nasir20241} investigates a secure cell-free \ac{ISAC} system, where distributed access points collaboratively serve \acp{CU} and perform target sensing in the presence of multiple eavesdroppers. For that, the authors propose to jointly optimize the communication and sensing beamforming vectors, with the goal of maximizing both the secrecy rate for all \acp{CU} and the sensing \ac{SNR} of the target.
Also, the work in \cite{10018908} studies outage beamforming strategies for \ac{DFRC} \ac{ISAC} systems, but does not touch upon the security vulnerabilities that are inherent to \ac{ISAC} systems.

Recently, \ac{FD} communications have emerged as a key enabler for \ac{ISAC} applications, thanks to their ability to support simultaneous \ac{DL} and \ac{UL} transmissions across the entire frequency band.
One of the main challenges in establishing in-band monostatic \ac{FD} communication is the detrimental effect of \ac{SI}. It constitutes a significant part of the received signal, drains high unnecessary power (up to $100\dB$ higher than those of the signals of interest \cite{10463523} \cite{10258345}), yet does not contain any useful information. There are three main components of \ac{SI}: leakage \ac{SI}, direct \ac{SI} or spillover, and reflected \ac{SI} \cite{10258345}. The major portion of \ac{SI} must be canceled in the analog domain, i.e. before it reaches analog to digital conversion, after which the residual \ac{SI} is treated in the digital domain.
From \ac{FD}-\ac{ISAC} perspective, the work in \cite{9724187} proposes an \ac{FD} waveform design for a monostatic \ac{ISAC} setup, wherein a single \ac{ISAC} node aims at simultaneously sensing a radar target while communicating with a communication receiver. The design can enhance the communication and the probability of target detection, assuming effective suppression of \ac{SI}, by time-multiplexing the classic pulsed radar waveform with \ac{SAC} signals. Furthermore, \cite{10158711} considers an \ac{FD}-\ac{ISAC} system, in which the \ac{BS} simultaneously executes target detection and communicates with multiple \ac{DL} and \ac{UL} users by reusing the same time and frequency resources, where the aim is to maximize the sum rate while minimizing power consumption. The authors in \cite{10274660} consider the joint active and passive beamforming design problem in RIS-assisted \ac{FD} \ac{UL} communication systems. 
The study in \cite{10159012} proposes that \ac{FD}-\ac{ISAC} can address the \textit{“echo-miss problem”}, a phenomenon where radar returns are suppressed by strong residual self-interference (SI) from the transmitter in a monostatic ISAC receiver. The solution involves jointly designing the beamformers, \ac{UL} transmit precoder, and \ac{DL} receive combiner to maximize communication rates, enhancing radar beampattern power, and suppressing residual \ac{SI}.
The authors in \cite{10221890} present a novel system termed \ac{ISABC}, which consists of an \ac{FD} \ac{BS} that extracts environmental information from its transmitted signal, a backscatter tag that reflects the signal, and a user who receives the data.
The authors discover that, while the power allocation at the \ac{BS} plays a crucial role in influencing user and sensing rates, it does not affect the backscatter rate. 
The work in \cite{9363029} emphasizes the design challenges in integrating communication and sensing/radar systems and investigates possible solutions, such as simultaneous transmission and reception, multibeam beamforming, and joint waveform optimization.
In addition, \cite{Allu20241} investigates an \ac{FD}-\ac{ISAC} system where the \ac{BS} is tasked with target detection while sharing the same resources for simultaneous \ac{DL}/\ac{UL} communications, where the focus is on tackling the robust energy efficiency maximization problem for a sustainable \ac{FD}-\ac{ISAC} architecture.
Also, \cite{Galappaththige20241} focuses on near-field \ac{FD}-\ac{ISAC} beamforming for multi-target detection.
Our previous work in \cite{10373185} addresses the problem of satisfying \ac{UL} and \ac{DL} secrecy rates, under \ac{ISMR} constraints, while minimizing the total power consumption of the \ac{FD} system, in the presence of eavesdropping targets.

The works in \cite{10323118}, \cite{10239482}, \cite{10478709}, and \cite{10443616} all investigate the application of \ac{RIS} in \ac{PLS} of \ac{ISAC} systems. The authors in \cite{10239482} use high-power radar signal as the interference to disturb the eavesdropper’s interception and \ac{RIS} is used to expand the coverage area. The work in \cite{10478709} considers \ac{RIS}-based backscatter systems, where an \ac{ISAC} \ac{BS} attempts to communicate with multiple \ac{IoT} devices in the presence of an aerial eavesdropper. Finally, the work in \cite{10443616} examines a \ac{RIS}-based \ac{THz} \ac{ISAC} system with delay alignment modulation. 
Also, \cite{10508296} utilizes \ac{RIS} for secure transmissions within \ac{UAV} \ac{ISAC} systems, whereby a methodology is proposed taking into account the velocity and positions of the \ac{UAV}.

\subsection{Contributions and Insights}
This work centers on a secure \ac{FD} design, where the \ac{DFRC} \ac{BS} transmits \ac{ISAC} signals optimized for \ac{SAC}, with the goal of preventing multiple eavesdropping targets to decode these signals in \ac{FD}-\ac{ISAC} mode.
To that purpose, we have summarized our contributions as follows.
\begin{itemize}
    \item \textbf{Maximum Secrecy Optimization for Secure \ac{FD} ISAC.} We first introduce our \ac{FD}-\ac{ISAC} scenario that includes \ac{DL} and \ac{UL} \acp{CU} associated with a a \ac{DFRC} \ac{BS} in the presence of multiple eavesdropping targets, where the \ac{DFRC} \ac{BS} would like to sense these eavesdropping targets. The injected \ac{AN} signal is not only used for securing the signal, but also aids in the sensing tasks through proper \ac{ISMR} regulations. We then formulate an optimization framework that aims at maximizing the \ac{FD} sum secrecy rate, which includes both \ac{UL} and \ac{DL} sum secrecy rates. For sensing, we impose an \ac{ISMR} constraint which also has the ability of revealing minimal communication information, while maximizing the radar backscattered signal for sensing purposes.
    \item \textbf{Solution via Taylor series approximations and block cyclic coordinate descent.} As the proposed maximum \ac{FD} secrecy optimization problem is non-convex and highly non-linear, we prove a series of lemmas, which are crucial as they allow us to transform the problem into a more tractable form. 
    Subsequently, we derive a method termed \ac{IJTB}, which cycles between two sub-problems, one of which produces the \ac{UL} beamformers in closed-form, while the other approximate problem, generates the \ac{UL} power allocation vector, the \ac{AN} covariance generator and the \ac{DL} beamforming vectors, thanks to a series of Taylor approximations applied to \textit{"convexify"} the problem. The proposed \ac{IJTB} algorithm for maximum-secrecy \ac{FD}-\ac{ISAC} iterates between the two problems and converges to a stable solution providing the optimal beamformers, artificial noise statistics, and power allocation vector for the \ac{UL} \acp{CU}, in addition to the \ac{UL} beamformers.
    \item \textbf{Extensive simulation results}. We present extensive simulation results that highlight the superiority, as well as the potential of the proposed design and algorithm with respect to many benchmarks, in terms of \ac{UL} and \ac{DL} sum secrecy rate, achieved \ac{ISMR}, in addition to algorithmic convergence.
    \end{itemize}
Unlike our previous work \cite{10373185}, this work essentially focuses on the maximization of the sum secrecy rate of both \ac{UL} and \ac{DL} users, under sensing and various power budget constraints. \textit{In particular, we are interested in studying the maximum sum secrecy rates for both \ac{UL} and \ac{DL}, under sensing and power constraints, which serves a complementary study to our previous work in \cite{10373185}, that mainly focuses on power efficient ways to satisfy secure and \ac{ISAC} constraints.}
Moreover, we unveil some important insights, i.e.
\begin{itemize}
	\item The \ac{IJTB} method consistently outperforms all benchmarks in terms of uplink (UL) sum secrecy rate, regardless of the distance of the eavesdropper or the \ac{UL} radial distance. More specifically, as the UL radial distance increases, the UL sum secrecy rate decreases for all methods. Meanwhile, \ac{IJTB} can adapt to different eavesdropper distances, maintaining a higher UL sum secrecy rate even as the eavesdropper gets closer. For instance, all benchmarks show a significant drop in the \ac{UL} sum secrecy rate, nearing zero when the eavesdropper distance reaches $17\meters$, highlighting their vulnerability when compared to \ac{IJTB}.
	\item Regarding \ac{AN} exploitation, \ac{IJTB} adapts well to varying eavesdropper distances, maintaining its performance consistently and outperforming all other methods. For example, when the \ac{UL} \ac{CU} is situated at $50\meters$, \ac{IJTB} achieves a sum secrecy rate of about $21.4 \bpspHz$, compared to $18.86 \bpspHz$ by the isotropic no-\ac{AN} benchmark. This shows that \ac{IJTB} can explore the extra \ac{DoF} offered by \ac{AN}, even when performing sensing tasks, i.e. under \ac{ISMR} constraints.
	\item From sensing perspective for \ac{ISAC}, when \ac{ISMR} constraint is stringent, the sum secrecy rate of IJTB drops below that of isotropic benchmarks, due to power allocation strategies that focus on minimizing sidelobe power. Despite this "drop", IJTB still achieves competitive sum secrecy rates, especially as the ISMR constraint becomes less restrictive, reflecting a favorable increase for sensing performance. Putting things in context, when \ac{ISMR} is tuned to $-20\dB$ (i.e. suggesting very pointy radar beams), \ac{IJTB} can still achieve a total sum secrecy rate of about $8.84 \bpspHz$, which is comparable with other benchmarks.
	\item \ac{IJTB} demonstrates rapid convergence, stabilizing in about $10$ iterations.
	\end{itemize}

\subsection{Organization \& Notation}
The detailed structure of the following paper is given as follows: The full-duplex integrated sensing and communication system model along with its key performance indicators is presented in Section II. The formulation of the proposed optimization framework bespoke for the secure \ac{FD}-\ac{ISAC} design problem is outlined in Section III. Consequently, its solution in a form of an algorithm is described in Section IV, while Section V describes the simulation findings. Lastly, the conclusion of the paper can be found in Section VI.

\textit{Notation:} A vector is denoted by a lower-case boldface letter, e.g. $\pmb{x}$, while its $\ell_2$ norm is denoted as $\Vert \pmb{x} \Vert$. The zero-vector is denoted by $\pmb{0}$, while the all-ones vector of appropriate dimensions is denoted by $\pmb{1}$. A matrix is denoted by an upper-case boldface letter and its transpose, conjugate, and transpose-conjugate are denoted by $(.)^T$, $(.)^*$, and $(.)^H$, respectively. A vector with all non-negative entries is denoted by $\pmb{x} \succeq \pmb{0}$, while a positive semi-definite matrix is denoted by $\pmb{A} \succeq \pmb{0}$. The $N \times N$ identity matrix is $\pmb{I}_N$. When it comes to matrix indexing, $[\pmb{A}]_{i,j}$ denotes the $(i,j)^{th}$ entry of matrix $\pmb{A}$ and $[\pmb{A}]_{:,j}$ denotes its $j^{th}$ column. The operator $\EVD$ stands for eigenvalue decomposition. The magnitude and the angle of any complex number $z\in \mathbb{C}$ are denoted by $\vert z \vert$ and $\angle z$, respectively. $\mathbb{E}\lbrace \rbrace$ denotes the statistical expectation. 


%% file: sections/system-model.tex
\begin{figure}[!t]
\centering
\includegraphics[width=3.5in]{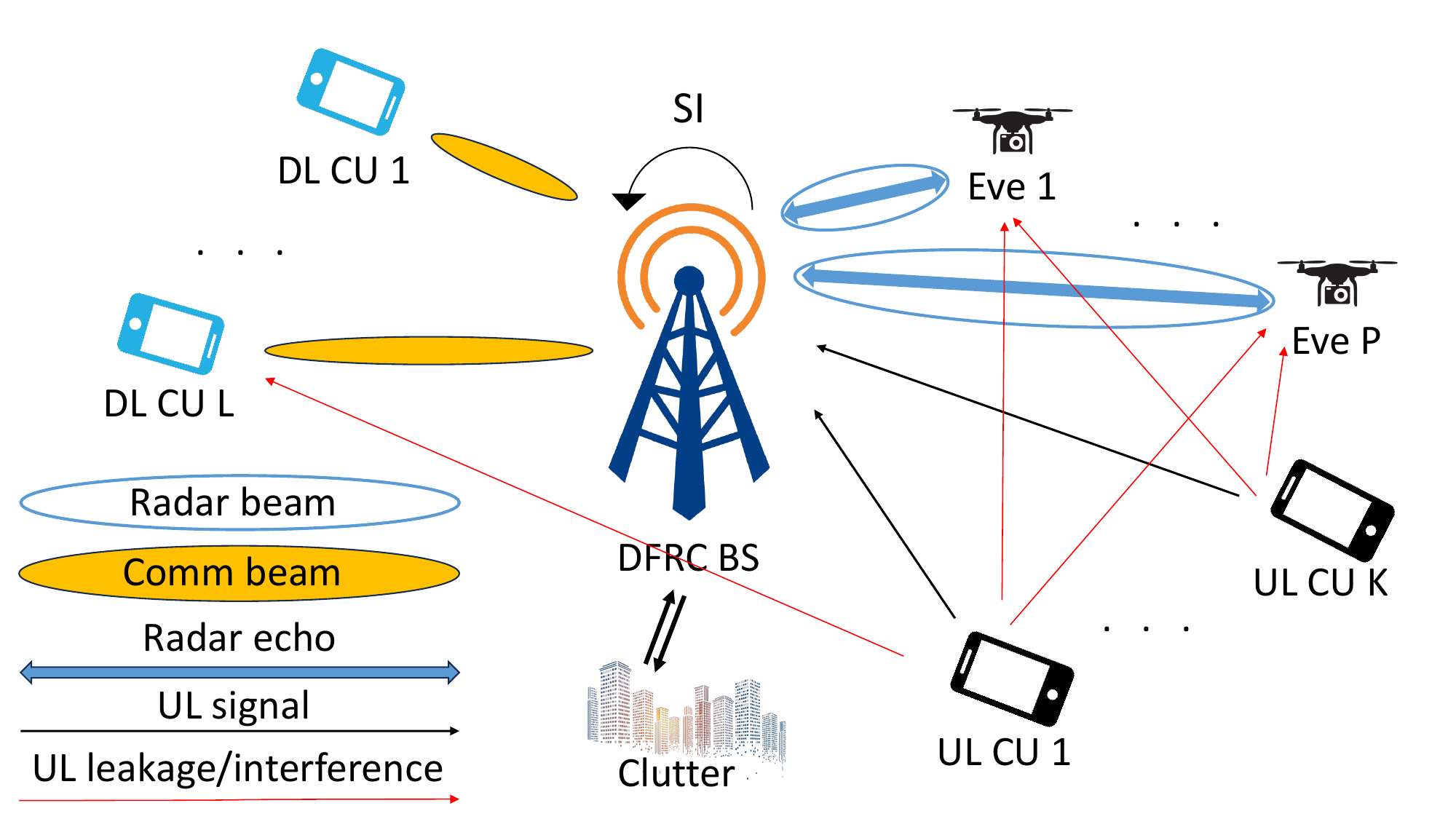}
\caption{The \ac{FD}-\ac{ISAC} system model which entails multiple malicious targets attempting to intercept \ac{UL} and \ac{DL} communications between the \ac{DFRC} \ac{BS} and the legitimate \ac{UL}/\ac{DL} \acp{CU}.}
\label{fig_1}
\end{figure}
\subsection{Full Duplex Radar and Communication Model}
\label{sec:FD-DFRC-model}
We assume an \ac{FD}-\ac{ISAC} scenario, in which there is a threat of malicious eavesdroppers intercepting data, as depicted in Fig.\ref{fig_1}. The DFRC BS considered is equipped with $N_T$ transmitting antennas and $N_R$ receiving antennas displaced via a mono-static radar setting with no isolation between both arrays. The following $N_T \times 1$ secure ISAC baseband signal vector is transmitted by the \ac{DFRC} \ac{BS}
\useshortskip
\begin{equation}
    \label{eq:signal-vector}
    \pmb{x} = \pmb{Vs}+\pmb{w},
\end{equation}
where $\pmb{V} \in \mathbf{C}^{N_T \times L}$ is the DFRC beamforming matrix with its $\ell ^{th}$ column loaded with the beamforming vector designated towards the $\ell ^{th}$ \ac{CU}; and $L$ denotes the total number of \ac{DL} \acp{CU}. Also, $\pmb{s}$ is a vector of transmit data symbols, whose $\ell ^{th}$ entry is the data symbol intended to the $\ell ^{th}$ \ac{DL} \ac{CU}. We assume independent symbols with unit variance, i.e $\mathbb{E}(\pmb{s}\pmb{s}^H) = \pmb{I}_L$. Moreover, a superimposed \ac{ISAC} \ac{AN} vector $\pmb{w}$, assumed to be independent from the signal $\pmb{s}$, is transmitted based on $\pmb{w} \sim \mathcal{N}(\pmb{0},\pmb{W})$, 
where $\pmb{W} \succeq \pmb{0}$ stands for the spatial \ac{ISAC} \ac{AN} covariance matrix, which is unknown, and has to be optimized for. Likewise, $\pmb{w}$ and $\pmb{s}$ are assumed to be independent of one another, in which case, $\trace(\pmb{W})$ would reflect the total transmit power invested on \ac{AN}. The beamformer $\pmb{V}$ is designed for DFRC \ac{ISAC} beamforming to maintain good \ac{SAC} properties. 
As a consequence, although the secure \ac{ISAC} signal $\pmb{x}$ is intended for the \ac{DL} \acp{CU}, it is also designed to function as a radar signal through both $\pmb{V}$ and $\pmb{W}$.

According to the ISAC design, this model does not distinctly separate \ac{SAC} signals, treating them as a unified signal instead. Consequently, the single-antenna DL \acp{CU} receive the following signal:
\useshortskip
\begin{equation}
 	\label{eq:DL-signal-2}
		{y}^{\mathrm{DL}}_\ell
	=
	\underbrace{\pmb{h}_{r,\ell}^H
	\pmb{v}_\ell s_\ell}_{\rm{desired}}
	+
	\underbrace{\sum\limits_{k=1}^K
	q_{k,\ell}v_k}_{\substack{\rm{UL-to-DL}\\\rm{interference} }}
	+
	\underbrace{\sum\limits_{\substack{\ell^{'} \neq \ell}}^L
	\pmb{h}_{r,\ell}^H
	\pmb{v}_{\ell^{'}} s_{\ell^{'}}}_{\substack{\rm{MU}\\\rm{interference} }}
	+
	\underbrace{\pmb{h}_{r,\ell}^H
	\pmb{w}}_{\substack{\rm{AN}\\\rm{interference} }}
	+
	n_\ell,
\end{equation}
where the desired and interfering terms are elucidated, and the last term, i.e.
$n_\ell$, is zero-mean \ac{AWGN} noise with variance $\sigma_{\ell}^2$, i.e. $n_\ell \sim \mathcal{N}(0,\sigma_{\ell}^2)$. 
Moreover, $\pmb{h}_{r,\ell}$ denotes the mmWave channel between the $\ell^{th}$ \ac{DL} \ac{CU} and the DFRC BS modeled as \cite{10229204} 
\useshortskip
\begin{equation*}
	\pmb{h}_{r,\ell}
=
\sqrt{\frac{\kappa_\ell}{\kappa_\ell + 1}}
\pmb{h}_{r,\ell}^{\mathrm{LoS}}
+
\sqrt{\frac{1}{\kappa_\ell + 1}}
\pmb{h}_{r,\ell}^{\mathrm{NLoS}}
\in 
\mathbb{C}^{N_T \times 1}
,
\end{equation*}
where $\kappa_\ell$ is the Rician $K$-factor of the $\ell^{th}$ user. Further, $\pmb{h}_{r,\ell}^{\mathrm{LoS}}$ is the LoS component between the $\ell^{th}$ user and the DFRC BS, and the \ac{NLoS} component consists of $N_{\mathrm{cl}}$ clusters as
\useshortskip
$
	\pmb{h}_{r,\ell}^{\mathrm{NLoS}}  =
\sqrt{1 \Big/ \sum_{q=1}^{N_{\mathrm{cl}}} N_{q}} \cdot \sum_{q=1}^{N_{\mathrm{cl}}} \sum_{r=1}^{N_{q}} u_{q,r} 
\pmb{a}_{N_T}(\varphi_{q,r})
$  \cite{7400949},
where $N_q$ is the number of propagation paths within the $q^{th}$ clusters.
The steering vector departing from \ac{AoD} $\varphi$ is denoted by $\pmb{a}_{N_T}(\varphi) \in \mathbb{C}^{N_T \times 1}$.
Additionally, we define $u_{q,r}$ as the path attenuation and $\varphi_{q,r}$ as \acp{AoD} of the $r^{th}$ propagation path within the $q^{th}$ cluster. As the number of clusters grows, the path attenuation coefficients and angles between users and the DFRC BS become randomly distributed. 
Additionally, $q_{k,\ell}$ represents the channel coefficient between the $k^{th}$ \ac{UL} \ac{CU} and the $\ell^{th}$ \ac{DL} user, which captures the \ac{UL}-to-\ac{DL} interference, and can be harmful when UL and DL users are in close proximity to each other. Furthermore, the $k^{th}$ \ac{UL} symbol, formed by the $k^{th}$ user and intended for the DFRC BS, is denoted as $v_k \in \mathbb{C}$, with power level $\mathbb{E}(\vert v_k \vert^2) = p_k, \forall k$. Due to the presence of clutter, a portion of the \ac{DL} signal scatters from the clutter towards the \ac{DL} \acp{CU}, which is contained within the \ac{DL} channels $\pmb{h}_{r,\ell}$.
The eavesdroppers are also intercepting \ac{DL} signals, so the signal at the $p^{th}$ eavesdroppers is
\begin{equation}
	\label{eq:pth-eve}
	y_p^{\mathrm{Eve}}
	=
	\sum\nolimits_{k=1}^K
	g_{k,p}v_k
	+
	\alpha_p
	\pmb{a}_{N_T}^H(\theta_p)
	\pmb{x}
	+
	z_p,
\end{equation} 
where $g_{k,p}$ is the complex channel coefficient between the $k^{th}$ \ac{UL} \ac{CU} and the $p^{th}$ eavesdropper. The path-loss complex coefficient between the DFRC BS and the $p^{th}$ eavesdropper is denoted by $\alpha_p$.
The steering vector arriving at \ac{AoA} $\theta$ is denoted by $\pmb{a}_{N_R}(\theta) \in \mathbb{C}^{N_R \times 1}$.
The angles of arrival (AoAs) for the eavesdroppers are denoted as $\theta_p$.
The AWGN at the $p^{th}$ eavesdropper is $z_p$, modeled as zero-mean with variance $\sigma_p^2$, i.e., $z_p \sim \mathcal{N}(0,\sigma_p^2)$.
We emphasize the relationship between the received signal at the $ \ell ^{th}$ legitimate DL user (as in \eqref{eq:DL-signal-2}) and the signal at the eavesdropper (as in \eqref{eq:pth-eve}). Both legitimate users and eavesdroppers receive the UL signals and the DFRC ISAC signal $ \pmb{x} $. However, the channel models differ between DFRC-DL (i.e., $ \pmb{h}_{r,\ell} $) and DFRC-eavesdropper (i.e., $ \pmb{a}(\theta_p) $). Eavesdroppers are considered radar targets with a strong dominant line-of-sight (LoS) component (a Rician channel with a large $ K $-factor) for localization and tracking applications. This model is particularly applicable in scenarios involving \acp{UAV} \cite{9737357}.

Specifically, the DFRC aims to infer additional sensing information about eavesdropping targets, such as delay and Doppler, without exposing the communication information embedded within the transmitted ISAC signal $ \pmb{x} $. Meanwhile, a set of $ K $ UL \acp{CU} associated with the DFRC base station (BS) coexist. It is important to note that while the DFRC BS transmits a secure signal vector $ \pmb{x} $, it also collects the UL signals along with other components clarified in this section. Let $ \pmb{h}_{t,k} $ be the UL channel vector between the $ k^{th} $ UL \ac{CU} and the DFRC BS. For Rayleigh fading, the UL channels $ \pmb{h}_{t,k} $ account for environmental clutter, where part of the signal propagates towards the clutter and back to the DFRC BS. Thus, the signal received by the DFRC BS from the single-antenna UL users is $ \sum\nolimits_{k=1}^K \pmb{h}_{t,k} v_k $.

We summarize the system model in Fig. (reference), which illustrates a group of UL and DL legitimate \acp{CU} associated with a DFRC BS. Additionally, a group of eavesdroppers intercepts the UL and DL signal exchanges. Note that the DL signal is an ISAC signal, and the DFRC BS directs beams towards the eavesdroppers to acquire physical sensing information about them. 
Furthermore, due to full-duplex (FD) operation, the simultaneous transmission of the signal vector $ \pmb{x} $ \textit{"leaks"} into the receiving unit of the DFRC BS, resulting in loop interference. This phenomenon, known as \ac{SI}, is a crucial component when targeting FD behavior, especially when the transmitting and receiving units of the DFRC BS are colocated without isolation. The receiving unit experiences an \ac{SI} component of $ \sqrt{\beta} \HSI \in \mathbb{C}^{N_R \times N_T} $, where $ \beta $ represents the residual \ac{SI} power. Under the direct-coupling channel model for \ac{SI}, the small-scale \ac{SI} fading channel reads $[\HSI]_{i,j}
	=
	\exp \left( j 2 \pi \frac{d_{i,j}}{\lambda} \right)$, which contains the phases introduced between each pair of transmitting and receiving antennas. Also, $\lambda$ is the wavelength and $ d_{i,j} $ is the distance between the $ i $-th receive and $ j^{th} $ transmit antennas.
Additionally, the DFRC BS aims to direct sensing power towards $ P $ eavesdroppers so that the  backscattered echo returns experience good sensing \ac{SNR}. To this end, the \textit{echo-plus-clutter} returns can be written as
\useshortskip
\begin{equation}
	\label{eq:echo}
	\pmb{y}_{\rm{e}+\rm{c}} = 
	\sum\nolimits_{p=1}^P
	\gamma_p \pmb{a}_{N_R}(\theta_p)\pmb{a}_{N_T}^H(\theta_p)\pmb{x} + \pmb{c}.
\end{equation} 
\useshortskip
In \eqref{eq:echo}, we denote $\gamma_p$ by the complex amplitude due to the $p^{th}$ eavesdropper and $\pmb{c} \sim \mathcal{N}(\pmb{0},\pmb{R}_c)$ is the clutter including the reflected signal from the \ac{UL}/\ac{DL} \acp{CU}. We assume the clutter covariance matrix $\pmb{R}_c \in \mathbb{C}^{N_R \times N_R}$, capturing all environmental clutter effects, to be constant and that the eavesdropper \acp{AoA} are known.
Due to the collocated transmit and receiving units, the \acp{AoA} and \acp{AoD} in equation \eqref{eq:echo} are identical. Lastly, the received signal at the DFRC BS is
\begin{equation}
	\label{eq:BS-signal}
	\pmb{y}^{\mathrm{DFRC}}
	=
	\sum\nolimits_{k=1}^K \pmb{h}_{t,k} v_k
	+
	\pmb{y}_{\rm{e}+\rm{c}}
	+
	\sqrt{\beta}\HSI \pmb{x}
	+
	\pmb{n}.
\end{equation}
In summary, the first part in \eqref{eq:BS-signal} is the signal due to the $K$ \ac{UL} \acp{CU}. 
The second part is the sensing \textit{echo-plus-clutter} returns. The third part is the \ac{SI}, and the \ac{AWGN} follows $\pmb{n} \sim \mathcal{N}(\pmb{0},\sigma_n^2 \pmb{I}_{N_R})$.
\subsection{Secrecy and sensing metrics}
Before tackling the secure \ac{FD}-\ac{ISAC} problem, it is important to define \acp{KPI} that will be critical for evaluating the system's overall performance and for developing an appropriate optimization framework.
\subsubsection{Secrecy rates}
First, we define the \ac{SINR} of the $\ell^{th}$ \ac{DL} \ac{CU}. Based on \eqref{eq:DL-signal-2}, and under the assumption of independent symbols, the \ac{SINR} for detecting the $\ell^{th}$ desired symbol, $s_{\ell}$, can be verified to be
\useshortskip
\begin{equation}
	\label{eq:SINR_DL}
	\Gamma_{\ell}^{\mathrm{DL}}
	= 
	\frac{ \pmb{h}_{r,\ell}^H\pmb{V}_\ell\pmb{h}_{r,\ell} } 
	     {\sum\limits_{\substack{k = 1  }}^K p_k \vert q_{k,\ell} \vert^2 +\sum\limits_{\substack{\ell^{'} \neq \ell}}^L \pmb{h}_{r,\ell}^H\pmb{V}_{\ell^{'}}\pmb{h}_{r,\ell} + \pmb{h}_{r,\ell}^H\pmb{W}\pmb{h}_{r,\ell}  +\sigma_{\ell}^2}, 
\end{equation}
where $\pmb{V}_{\ell} = \pmb{v}_{\ell} \pmb{v}^H_{\ell}$. Moreover, the achievable rate of transmission of the $\ell^{th}$ \ac{DL} \ac{CU} is
\useshortskip
\begin{equation}
\label{eq:rate-equation}
	R_{\ell}^{\mathrm{DL}} = \log_2( 1 + \Gamma_{\ell}^{\mathrm{DL}} ), \quad \forall \ell = 1 \ldots L
\end{equation}
To evaluate \ac{UL} communication performance at the DFRC BS, we also define an \ac{UL} \ac{SINR}. Assuming the \ac{BS} applies $K$ beamforming vectors $\pmb{u}_1 \ldots \pmb{u}_K$ onto the receive signal in \eqref{eq:BS-signal}, we can then define a receive \ac{SINR} quantifying the performance between the $k^{th}$ \ac{UL} \ac{CU} and the \ac{DFRC} \ac{BS} intending to detect $v_k$, which can be shown to be
\useshortskip
\begin{equation}
	\label{eq:SINR_UL1}
	\Gamma_{k}^{\mathrm{UL}}
	=
	\frac
	{ p_k\pmb{u}_k^H \pmb{h}_{t,k}\pmb{h}^H_{t,k} \pmb{u}_k }
	{\pmb{u}_k^H\pmb{\Phi}_k(\pmb{Q},\pmb{p})\pmb{u}_k},
\end{equation}
where
\useshortskip
\begin{equation}
	\label{eq:phi_k_eqn}
	\pmb{\Phi}_k(\pmb{Q},\pmb{p}) = \sum\nolimits_{\substack{ k' \neq k }}^K  p_{k'}\pmb{h}_{t,k'}\pmb{h}^H_{t,k'}   +
	 \pmb{C}\pmb{Q}\pmb{C}^H + 
	 \pmb{R}_c +
	\sigma_n^2 \pmb{I}
\end{equation}
and $\pmb{C} = \sum\nolimits_{p=1}^P \gamma_p \pmb{a}_{N_R}(\theta_p)\pmb{a}_{N_T}^H(\theta_p) + \sqrt{\beta}\HSI $.
Also, $\pmb{Q} = \pmb{V}\pmb{V}^H + \pmb{W}$. Similarly, we can define the achievable rate of transmission of the $k^{th}$ \ac{UL} \ac{CU} as
\useshortskip
\begin{equation}
\label{eq:rate-equation-2}
	R_{k}^{\mathrm{UL}} = \log_2( 1 + \Gamma_{k}^{\mathrm{UL}} ), \quad \forall k = 1 \ldots K
\end{equation}
Prior to expressing the \ac{UL} and \ac{DL} \textit{secrecy rates}, it would be appropriate to define the sum-rates per eavesdropper. We start by defining a receive \ac{SINR} between the $p^{th}$ eavesdropper and the $k^{th}$ \ac{UL} \ac{CU}, who aims at detecting $v_k$. After some operations, it can be proven that the corresponding \ac{SINR} reads
\useshortskip
\begin{equation}
	\Gamma_{p,k}^{\mathrm{Eve}}
	=
	\frac
	{ p_k \vert g_{k,p} \vert^2  }
	{\sum\limits_{\substack{ k' \neq k }}^K p_{k'} \vert g_{k',p} \vert^2 + \vert \alpha_p \vert^2 \pmb{a}_{N_T}^H(\theta_p)\pmb{Q}\pmb{a}_{N_T}(\theta_p) + \sigma_p^2}.
\end{equation}
Similarly, we can express the receive \ac{SINR} between the $p^{th}$ eavesdropper and the \ac{DFRC} \ac{BS},
\useshortskip
\begin{equation}
	\Gamma_{p,\mathrm{DFRC}}^{\mathrm{Eve}}
	=
	\frac
	{\vert \alpha_p \vert^2 \pmb{a}_{N_T}^H(\theta_p)\pmb{VV}^H\pmb{a}_{N_T}(\theta_p)}
	{\sum\limits_{\substack{k = 1  }}^K p_k \vert g_{k,p} \vert^2 + 
	\vert \alpha_p \vert^2 \pmb{a}_{N_T}^H(\theta_p)\pmb{W}\pmb{a}_{N_T}(\theta_p)
	 + \sigma_p^2 },
\end{equation}
Having the \ac{DL} and \ac{UL} \ac{SINR} expressions per eavesdropper defined, we will now define the \ac{DL} sum secrecy rate following \cite{7086319} as follows
\useshortskip
\begin{equation}
\label{eq:sum-secrecy-rate-DL}
	\SR_{\mathrm{DL}}
	=
	\sum\nolimits_\ell
	\log_2(1+\Gamma_{\ell}^{\mathrm{DL}})
	-
	\sum\nolimits_p
	\log_2(1+\Gamma_{p,\mathrm{DFRC}}^{\mathrm{Eve}})
\end{equation}
Likewise, the \ac{UL} sum secrecy rate reads
\useshortskip
\begin{equation}
\label{eq:sum-secrecy-rate-UL}
\SR_{\mathrm{UL}}
	=\sum\nolimits_{k}
	\log_2(1+\Gamma_{k}^{\mathrm{UL}})
	-
	\sum\nolimits_{p,k}
	\log_2(1+\Gamma_{p,k}^{\mathrm{Eve}}).
\end{equation}
\subsubsection{Sensing metrics}
The DFRC BS seeks to manage beams using each of its transmit vectors, $\pmb{x}$, to ensure highly directional radar transmissions aimed at the $P$ eavesdroppers. This simultaneous targeting allows the DFRC BS to extract wireless environmental features, using the backscattered echoes, and analyze the physical attributes of each eavesdropper e.g. delay-Doppler profile. For this, we utilize the \ac{ISMR} \ac{KPI} to assess radar performance based on the transmit beampattern. This ratio quantifies the energy in the sidelobes relative to the mainlobe as given below
\useshortskip
\begin{equation}
\label{eq:ISMR_def}
	\ISMR \triangleq \frac
	{\int\nolimits_{\pmb{\Theta}_s}\pmb{a}_{N_T}^H(\theta) \pmb{R}_{\pmb{xx}} \pmb{a}_{N_T}(\theta) \ d\theta }
	{\int\nolimits_{\pmb{\Theta}_m}\pmb{a}_{N_T}^H(\theta) \pmb{R}_{\pmb{xx}} \pmb{a}_{N_T}(\theta) \ d\theta }
=
\frac
	{ \trace\big( \pmb{R}_{\pmb{xx}} \pmb{A}_s \big) }
	{\trace\big( \pmb{R}_{\pmb{xx}} \pmb{A}_m \big)},
\end{equation}
where $\pmb{\Theta}_s$ and $\pmb{\Theta}_m$ are sets containing the directions toward the side-lobe and main-lobe components, respectively. Moreover, $\pmb{R}_{\pmb{xx}}$ represents the transmit covariance matrix, i.e, $\pmb{R}_{\pmb{xx}}
	=
	\pmb{V}\pmb{V}^H
	+
	\pmb{W}$.
The \ac{ISMR} could be reformulated after integrating over directions as per \eqref{eq:ISMR_def}, namely $\pmb{A}_{s /m} = \int\nolimits_{\pmb{\Theta}_{s/m}} \pmb{a}_{N_T}(\theta)\pmb{a}_{N_T}^H(\theta) \ d\theta  $. For this paper, $P$ mainlobes are chosen to be centered around $\theta_1 \ldots \theta_P$, respectively. As a consequence, the mainlobe is given as
\useshortskip
\begin{equation}
	\label{eq:Mainlobe-expression}
	\Theta_m = \bigcup\nolimits_{p=1}^{P} \big\lbrace [\theta_p - \theta_p^{\mathrm{low}}; \theta_p - \theta_p^{\mathrm{high}} ] \big\rbrace.
\end{equation}
where $\theta_p^{\mathrm{low}},\theta_p^{\mathrm{high}}$ define the lower and upper angular regions of the $p^{th}$ mainlobe.
The ISMR metric is designed to focus power in the mainlobe regions while minimizing power in the sidelobes to suppress unwanted echo bounces, such as clutter and signal-dependent interference. By optimizing the beams around the target AoAs and simultaneously attenuating clutter returns, the signal-clutter-noise ratio is improved.
The ISMR metric is adaptable for target localization and tracking by integrating sophisticated tracking methods to steer mainlobes, while also allowing for sidelobe power reduction to minimize unwanted returns, such as clutter. Additionally, ISMR offers flexibility in managing uncertainties in eavesdropper locations, aligning the necessity for focused energy on eavesdroppers with minimal communication leakage through appropriate \ac{AN} covariance and beamforming design.
%

%% file: sections/optimization-framework.tex
In this section, we propose a secure \ac{FD}-\ac{ISAC} sum secrecy rate maximization problem with the aim of maximizing the total sum secrecy rate of both \ac{UL} and \ac{DL} \acp{CU} given the power budget constraints for \ac{DL} beamforming, \ac{UL} power allocation, and AN generation. For this purpose, we formulate the following optimization problem
\useshortskip
\begin{equation}
 \label{eq:problem1}
\begin{aligned}
(\mathcal{P}_1):
\begin{cases}
\max\limits_{\lbrace \substack{ \pmb{p},
									 \pmb{W},  
									 \pmb{V},
                                  \pmb{U}
                                    }  \rbrace}&  \SR_{\mathrm{DL}} + \SR_{\mathrm{UL}} \\
\textrm{s.t.}
 &  \trace(\pmb{VV}^H) \leq P_{\rm{DL}}^V , \  \trace(\pmb{W}) \leq P_{\rm{DL}}^W,  \\
&  \sum\limits_{k=1}^K p_k \leq  P_{\rm{UL}}, \\
& \ISMR \leq    \ISMR_{\rm{max}} , \\
& \pmb{W} \succeq \pmb{0}, \ \pmb{p} \succeq \pmb{0}.   \\
\end{cases}
\end{aligned}
\end{equation}
The problem in $(\mathcal{P}_1)$ intends to maximize the overall \ac{UL} and \ac{DL} sum secrecy rates. The first set of constraint imposes a total power budget on \ac{DL} communications. More specifically, $P_{\rm{DL}}^V $ is the power budget for communication beamforming and $P_{\rm{DL}}^W$ is the power budget for \ac{AN} generation. 
Moreover, the second constraint imposes a power budget on the \ac{UL} \acp{CU}, whereby the overall \ac{UL} power budget is $P_{\rm{UL}}$.
The third constraint imposes the maximum \ac{ISMR} level, through $\ISMR_{\rm{max}} $ and the last constraint imposes positive semi-definite constraint on the covariance matrix $\pmb{W}$ and entry-wise positivity on the \ac{UL} power allocation vector.
It is worth pointing out that vector $\pmb{w}$ is not only being employed for \ac{AN} purposes but also to enhance the \ac{DoF} of the transmitted signal, hence improving sensing performance. In particular, the optimization process incorporates not only the beamformer $\pmb{V}$, but also the associated \ac{AN} covariance matrix $\pmb{W}$, to achieve the desired \ac{ISMR}. This positively impacts target detection and localization performance.
To this end, the problem in \eqref{eq:problem1} can be equivalently re-written as follows
\vspace{-0.4cm}
\useshortskip
\begin{equation}
 \label{eq:problem11}
\begin{aligned}
(\mathcal{P}_{1.1}):
\begin{cases}
\max\limits_{\lbrace \substack{ \pmb{p},
									 \pmb{W},  
									 \{\pmb{V}_\ell\}_{\ell=1}^L,
                                  \pmb{U}
                                    }  \rbrace}&  \SR_{\mathrm{DL}} + \SR_{\mathrm{UL}} \\
\textrm{s.t.}
 &  \sum\limits_{\ell=1}^L\trace(\pmb{V}_{\ell}) \leq P_{\rm{DL}}^V , \  \trace(\pmb{W}) \leq P_{\rm{DL}}^W , \\
&  \sum\limits_{k=1}^K p_k \leq  P_{\rm{UL}}, \\
& \ISMR \leq    \ISMR_{\rm{max}} , \\
& \pmb{W} \succeq \pmb{0}, \ \pmb{p} \succeq \pmb{0}, \pmb{V}_\ell \succeq \pmb{0}, \forall \ell,   \\
& \rank(\pmb{V}_\ell) = 1, \forall \ell.
\end{cases}
\end{aligned}
\end{equation}
The equivalent problem is non-convex due to the $\rank-1$ constraint. Hence, we propose to drop this constraint to arrive at
\vspace{-0.4cm}
\useshortskip
\begin{equation}
 \label{eq:problem12}
\begin{aligned}
(\mathcal{P}_{1.2}):
\begin{cases}
\max\limits_{\lbrace \substack{ \pmb{p},
									 \pmb{W},  
									 \{\pmb{V}_\ell\}_{\ell=1}^L,
                                  \pmb{U}
                                    }  \rbrace}&  \SR_{\mathrm{DL}} + \SR_{\mathrm{UL}} \\
\textrm{s.t.}
 &  \sum\limits_{\ell=1}^L\trace(\pmb{V}_{\ell}) \leq P_{\rm{DL}}^V , \  \trace(\pmb{W}) \leq P_{\rm{DL}}^W,  \\
&  \sum\limits_{k=1}^K p_k \leq  P_{\rm{UL}}, \\
& \ISMR \leq    \ISMR_{\rm{max}} , \\
& \pmb{W} \succeq \pmb{0}, \ \pmb{p} \succeq \pmb{0}, \pmb{V}_\ell \succeq \pmb{0}, \forall \ell.
\end{cases}
\end{aligned}
\end{equation}
Fortunately, all the constraints in the above problem are convex, however the problem is still non-convex due to the sum secrecy rate objective function. 
In the subsequent section, we address the challenge of non-convexity by introducing an algorithm designed to efficiently solve the \ac{FD}-\ac{ISAC} sum secrecy rate maximization problem.

%% file: sections/algorithmicderivation.tex
In this section, we propose IJTB, i.e. iterative joint Taylor-BCCD, as a potential algorithm that efficiently solves the \ac{FD}-\ac{ISAC} sum secrecy rate maximization problem formulated via $(\mathcal{P}_1)$ \eqref{eq:problem1} through its relaxed version $(\mathcal{P}_{1.2})$.
In order to design IJTB, it is crucial to convexify the problem $(\mathcal{P}_{1.2})$ so as to successfully iterate between the underlying parameters.
For the sake of compact representation, we define the interference received by the $\ell ^{\rm{th}}$ DL \ac{CU}, and the interference at the $p^{\rm{th}}$ eavesdropper, respectively, as follows,
\useshortskip
\begin{align}
	{\rm{I}}_{\ell}^{\mathrm{DL}} &= 
	\sum_{\substack{k }} p_k \vert q_{k,\ell} \vert^2 +\sum_{\substack{ \ell' \neq \ell}} \trace \left( \pmb{V}_{\ell'}\pmb{H}_{r,\ell} \right) +\trace \left(\pmb{W}\pmb{H}_{r,\ell} \right), \\
{\rm{I}}_{p, \mathrm{DFRC}}^{\mathrm{Eve}} &= 
	\sum_{\substack{k }} p_k \vert g_{k,p} \vert^2 + 
	\vert \alpha_p \vert^2 \trace \left(\pmb{W}\pmb{A}(\theta_p)\right).
\end{align}
The corresponding useful signal components of the $\ell ^{\rm{th}}$ DL \ac{CU} and the $p^{\rm{th}}$ eavesdropper are then ${\rm{S}}_{\ell}^{\mathrm{DL}} = \trace(\pmb{V}_{\ell}\pmb{H}_{r,\ell})$ and ${\rm{S}}_{p, \mathrm{DFRC}}^{\mathrm{Eve}} = \sum_\ell \trace \left(\vert \alpha_p \vert^2 \pmb{V}_\ell\pmb{A}(\theta_p)\right)$, respectively.
Therefore, the \ac{DL} sum secrecy rate can now be arranged as
\useshortskip
\begin{equation}
	\label{eq:SR_DL_rearranged}
 \begin{aligned}
 \SR_{\mathrm{DL}} &= 
 \sum_{\ell}\log _2\left({\rm{S}}_{\ell}^{\mathrm{DL}} + {\rm{I}}_{\ell}^{\mathrm{DL}} + \sigma_{\ell}^2 \right) \\ &+ \sum_{p} \log _2\left({\rm{I}}_{p, \mathrm{DFRC}}^{\mathrm{Eve}} + \sigma_p^2 \right) \\
 & - \underbrace{\sum_{\ell}\log _2\left({\rm{I}}_{\ell}^{\mathrm{DL}} +\sigma_{\ell}^2 \right)}_{\phi_1}  \\& -\underbrace{\sum_{p} \log _2\left({\rm{S}}_{p, \mathrm{DFRC}}^{\mathrm{Eve}} + {\rm{I}}_{p, \mathrm{DFRC}}^{\mathrm{Eve}} + \sigma_p^2\right)}_{\phi_2} 
 \end{aligned}
\end{equation}	
In this form, the \ac{DL} sum secrecy rate is not concave due to convexity of $-\phi_1$ and $-\phi_2$ terms. Therefore, we introduce the following lemmas with the intent of "convexifying" the problem.

\noindent\textit{Lemma 1:} The term $\phi_1$ can be approximated to an affine function by applying Taylor expansion around $(\widetilde{\pmb{p}}, \{\widetilde{\pmb{V}}_l\}_{l=1}^L, \widetilde{\pmb{W}})$
\useshortskip
\begin{equation}
\label{eq:phi1_final}
	\begin{aligned}
		\phi_1 & \approx \frac{1}{\ln 2}\sum_{\ell}\Bigg\{\ln\left|\widetilde{{\rm{I}}}_{\ell}^{\mathrm{DL}}+\sigma_{\ell}^2\right| \\
        &+ \dfrac{\sum_k \vert q_{k,\ell} \vert^2 {\Delta}_{p_k} + \operatorname{tr}\left(\pmb{H}_{r,\ell} \left[\sum_{s \neq \ell}\pmb{\Delta}_{\pmb{V}_s} + \pmb{\Delta}_{\pmb{W}}\right]\right)}{\widetilde{{\rm{I}}}_{\ell}^{\mathrm{DL}}+\sigma_{\ell}^2}\Bigg\},
	\end{aligned}
\end{equation}
where $\Delta _{p_k} = p_k - \widetilde{p}_k$, $\Delta _{\pmb{V}_s} = \pmb{V}_s - \widetilde{\pmb{V}}_s$, and $\Delta _{\pmb{W}} = \pmb{W} - \widetilde{\pmb{W}}$. In addition, 
\useshortskip
\begin{align*}
	\widetilde{\rm{I}}_{\ell}^{\mathrm{DL}} &= 
	\sum_{\substack{k }} \widetilde{p}_k \vert q_{k,\ell} \vert^2 +\sum_{\substack{ \ell' \neq \ell}} \trace \left( \widetilde{\pmb{V}}_{\ell'}\pmb{H}_{r,\ell} \right) +\trace \left(\widetilde{\pmb{W}}\pmb{H}_{r,\ell} \right).
\end{align*}
\textbf{Proof}: See Appendix \ref{sec:appendix1}.

Moreover, the function $\phi_2$ can also be approximated in a similar manner as $\phi_1$, which is detailed in the following lemma.

\noindent\textit{Lemma 2:} Similarly to Lemma 1, $\phi_2$ can also be approximated to an affine function by applying Taylor expansion around $(\widetilde{\pmb{p}}, \{\widetilde{\pmb{V}}_l\}_{l=1}^L, \widetilde{\pmb{W}})$:

\useshortskip
 \begin{equation}
 \label{eq:DL-phi2}
	\begin{aligned}
		 \phi_2 &\approx \frac{1}{\ln 2}\sum_{p}\Bigg\{\ln\vert \widetilde{{\rm{S}}}_{p, \mathrm{DFRC}}^{\mathrm{Eve}} + \widetilde{{\rm{I}}}_{p, \mathrm{DFRC}}^{\mathrm{Eve}} + \sigma_p^2\vert \\
        &+ \dfrac{\sum_k\vert g_{k,p} \vert^2{\Delta}_{p_k} +  \vert \alpha_p \vert^2\operatorname{tr}\left(\pmb{A}(\theta_p)\left[\sum_s\pmb{\Delta}_{\pmb{V}_s} + \pmb{\Delta}_{\pmb{W}}\right]\right)}{\widetilde{{\rm{S}}}_{p, \mathrm{DFRC}}^{\mathrm{Eve}} + \widetilde{{\rm{I}}}_{p, \mathrm{DFRC}}^{\mathrm{Eve}} + \sigma_p^2}\Bigg\},
	\end{aligned}
\end{equation}
where
\useshortskip
\begin{align}
\widetilde{\rm{I}}_{p, \mathrm{DFRC}}^{\mathrm{Eve}} &= 
	\sum_{\substack{k }} \widetilde{p}_k \vert g_{k,p} \vert^2 + 
	\vert \alpha_p \vert^2 \trace \left(\widetilde{\pmb{W}}\pmb{A}(\theta_p)\right), \\
	\widetilde{\rm{S}}_{p, \mathrm{DFRC}}^{\mathrm{Eve}} &= \sum_\ell \trace \left(\vert \alpha_p \vert^2 \widetilde{\pmb{V}}_\ell\pmb{A}(\theta_p)\right).
\end{align}
\textbf{Proof:} The proof follows similar steps as that of \textit{Lemma 1}.
Finally, the complete approximate expression for the \ac{DL} sum secrecy rate is given at the bottom of the page in equation \eqref{eq:SR_DL_bottom}, which is readily verified to be concave in $({\pmb{p}}, \{{\pmb{V}}_l\}_{l=1}^L, {\pmb{W}})$.
\useshortskip
 \begin{figure*}[b]
 \begin{equation}
	\label{eq:SR_DL_bottom}
 \begin{aligned}
 \wideparen{\SR}_{\mathrm{DL}} &= 
 \frac{1}{\ln2}\Bigg\{\sum_{\ell}\Bigg\{\ln\left({\rm{S}}_{\ell}^{\mathrm{DL}} + {\rm{I}}_{\ell}^{\mathrm{DL}} + \sigma_{\ell}^2 \right) - \ln\left|\widetilde{{\rm{I}}}_{\ell}^{\mathrm{DL}}+\sigma_{\ell}^2\right| - \dfrac{ \sum_k \vert q_{k,\ell} \vert^2 {\Delta}_{p_k} + \operatorname{tr}\left(\pmb{H}_{r,\ell} \left[\sum_{s \neq \ell}\pmb{\Delta}_{\pmb{V}_s} + \pmb{\Delta}_{\pmb{W}}\right]\right)}{\widetilde{{\rm{I}}}_{\ell}^{\mathrm{DL}}+\sigma_{\ell}^2}\Bigg\} \\
 & + \sum_{p} \left.\left\{\ln\left({\rm{I}}_{p, \mathrm{DFRC}}^{\mathrm{Eve}} + \sigma_p^2 \right) - \ln\vert \widetilde{{\rm{S}}}_{p, \mathrm{DFRC}}^{\mathrm{Eve}} + \widetilde{{\rm{I}}}_{p, \mathrm{DFRC}}^{\mathrm{Eve}} + \sigma_p^2\vert - \dfrac{\sum_k\vert g_{k,p} \vert^2{\Delta}_{p_k} + \vert \alpha_p \vert^2\operatorname{tr}\left(\pmb{A}(\theta_p)\left[\sum_s\pmb{\Delta}_{\pmb{V}_s} + \pmb{\Delta}_{\pmb{W}}\right]\right)}{\widetilde{{\rm{S}}}_{p, \mathrm{DFRC}}^{\mathrm{Eve}} + \widetilde{{\rm{I}}}_{p, \mathrm{DFRC}}^{\mathrm{Eve}} + \sigma_p^2}\right\}\right\}
 \end{aligned}
\end{equation}	
\end{figure*}
Next, we proceed to the \ac{UL} sum secrecy rate given in equation \eqref{eq:sum-secrecy-rate-UL}.
Note that $\Gamma_{k}^{\mathrm{UL}}$ forms a generalized Rayleigh quotient in $\pmb{u}_k$ as given in its form in \eqref{eq:SINR_UL1}. 
Also note that $\Gamma_{k}^{\mathrm{UL}}$ is the only quantity that depends on $\pmb{u}_k$. 
Based on this observation, the maximum of $\Gamma_{k}^{\mathrm{UL}}$ is attained by setting
\useshortskip
\begin{equation}
	\label{eq:closed-form-uk}
	\widehat{\pmb{u}}_k 
	=
	\pmb{\Phi}_k^{-1}(\pmb{Q},\pmb{p})
	 \pmb{h}_{t,k}, \quad \forall k,
\end{equation}
Using \eqref{eq:closed-form-uk} in \eqref{eq:SINR_UL1}, we get
\useshortskip
	\begin{equation}
	\widehat{\Gamma}_{k}^{\mathrm{UL}}
	= 
	 p_k
	 \pmb{h}^H_{t,k}
	\pmb{\Phi}_k^{-1}(\pmb{Q},\pmb{p}) 
	 \pmb{h}_{t,k}.
\end{equation}
Now that the \ac{UL} beamforming vectors have maximized the \ac{UL} \acp{SINR}, the maximization of the \ac{UL} sum secrecy rate can be performed based on the remaining variables, namely $({\pmb{p}}, \{{\pmb{V}}_l\}_{l=1}^L, {\pmb{W}})$, as the \ac{UL} sum secrecy rate at this point is
\useshortskip 
\begin{equation}
\label{eq:sum-secrecy-rate-UL}
\widehat{\SR}_{\mathrm{UL}}
	=\sum\nolimits_{k}
	\log_2(1+\widehat{\Gamma}_{k}^{\mathrm{UL}})
	-
	\sum\nolimits_{p,k}
	\log_2(1+\Gamma_{p,k}^{\mathrm{Eve}}),
\end{equation}
which is basically the \ac{UL} sum secrecy rate after \ac{UL} beamforming at the \ac{DFRC} \ac{BS}. On one hand, we have less variables to maximize the \ac{UL} sum secrecy rate, but on the other hand, this rate is still highly non-convex. To "convexify" it, we first lower-bound the first term as 
\useshortskip
\begin{equation}
	\sum\nolimits_{k}
	\log_2(1+\varpi_k)
	\leq 
	\sum\nolimits_{k}
	\log_2(1+\widehat{\Gamma}_{k}^{\mathrm{UL}}),
\end{equation} 
where we introduced $k$ slack variables satisfying $0 \leq \varpi_k \leq \widehat{\Gamma}_{k}^{\mathrm{UL}}$ for all $k$.
Next, a Taylor series expansion around $\widetilde{\boldsymbol{\Phi}}_k$ can be conducted, hence giving us a more suitable bound
\useshortskip
\begin{equation}
	\frac{\varpi_k}{p_k} \leq \pmb{h}_{t,k}^H\widetilde{\boldsymbol{\Phi}}_k^{-1} \pmb{h}_{t,k}-
\pmb{h}_{t,k}^H\widetilde{\boldsymbol{\Phi}}_k^{-1}\left(\boldsymbol{\Phi}_k-\widetilde{\boldsymbol{\Phi}}_k\right)\widetilde{\boldsymbol{\Phi}}_k^{-1} \pmb{h}_{t,k}, \forall k .
\end{equation}
However, this constraint is still non-convex due to it's fractional nature, i.e. $\frac{\varpi_k}{p_k} $ is not a convex function in its underlying variables. Therefore, we apply the following transformation through injecting slack variables $\vartheta_1 \ldots \vartheta_K$ appearing in quadratic structure, viz.
\useshortskip
\begin{equation}
	\left\{\begin{array}{l}
\frac{\vartheta_k^2}{p_k} \leq \pmb{h}_{t,k}^H\widetilde{\boldsymbol{\Phi}}_k^{-1} \pmb{h}_{t,k}-
\pmb{h}_{t,k}^H\widetilde{\boldsymbol{\Phi}}_k^{-1}\left(\boldsymbol{\Phi}_k-\widetilde{\boldsymbol{\Phi}}_k\right)\widetilde{\boldsymbol{\Phi}}_k^{-1} \pmb{h}_{t,k}, \quad \forall k . \\
\varpi_k \leq \vartheta_k^2,
\end{array}\right.
\end{equation}
The first constraint is easily verified to be convex, however the second constraint is non-convex (concave less than zero), but can be remedied through a first-order approximation around $\widetilde{\vartheta}_1 \ldots \widetilde{\vartheta}_K$ as follows
\useshortskip
\begin{equation}
\label{eq:set-of-constraints-for-bounding-the-UL-sum-secrecy-rate}
\left\{\begin{array}{l}
\frac{\vartheta_k^2}{p_k} \leq \pmb{h}_{t,k}^H\widetilde{\boldsymbol{\Phi}}_k^{-1} \pmb{h}_{t,k}-
\pmb{h}_{t,k}^H\widetilde{\boldsymbol{\Phi}}_k^{-1}\left(\boldsymbol{\Phi}_k-\widetilde{\boldsymbol{\Phi}}_k\right)\widetilde{\boldsymbol{\Phi}}_k^{-1} \pmb{h}_{t,k} \\
\varpi_k \leq \widetilde{\vartheta}_k^2+2 \widetilde{\vartheta}_k\left(\vartheta_k-\widetilde{\vartheta}_k\right),
\end{array} \forall k\right.
\end{equation}
This is equivalent to lower bounding the \ac{UL} sum secrecy rate under the set of constraints in \eqref{eq:set-of-constraints-for-bounding-the-UL-sum-secrecy-rate}
\begin{equation}
\label{eq:sum-secrecy-rate-UL-loweer-bound}
\sum\nolimits_{k}
\log_2(1+\varpi_k)
-
\sum\nolimits_{p,k}
\log_2(1+\Gamma_{p,k}^{\mathrm{Eve}})
\leq 
\widehat{\SR}_{\mathrm{UL}}
\end{equation}
Next, we proceed in convexifying the second term appearing in the lower bound of \eqref{eq:sum-secrecy-rate-UL-loweer-bound}. Indeed, we have that
\useshortskip
\begin{equation}
\label{eq:sum-secrecy-rate-UL-loweer-bound-alternative}
\begin{split}
\sum\limits_{k}
	\log_2(1+\varpi_k)
	&-\overbrace{\sum\limits_{p,k}
	\log_2({\rm{S}}_{p, k}^{\mathrm{Eve}} + {\rm{I}}_{p, k}^{\mathrm{Eve}} + \sigma_p^2)}^{K\phi_3}\\
    &+
    \sum\limits_{p,k}
	\log_2({\rm{I}}_{p, k}^{\mathrm{Eve}} + \sigma_p^2)
\leq 
\widehat{\SR}_{\mathrm{UL}}
\end{split}
\end{equation}
where ${\rm{S}}_{p, k}^{\mathrm{Eve}} = p_k \vert g_{k,p} \vert^2$ highlights the useful part of the received at the $p^{\rm{th}}$ eavesdropper and ${\rm{I}}_{p, k}^{\mathrm{Eve}} = \sum_{\substack{k'\neq k}}^K p_{k'} \vert g_{k',p} \vert^2 + \vert \alpha_p \vert^2 \pmb{a}_{N_T}^H(\theta_p)\pmb{Q}\pmb{a}_{N_T}(\theta_p)$ contains the interference the $p^{th}$ eavesdropper sees.
Note that although ${\rm{S}}_{p, k}^{\mathrm{Eve}}$ and ${\rm{I}}_{p, k}^{\mathrm{Eve}}$ each depend on $k$, their sum is independent of $k$. Thus, we will denote it by $\Psi _p^{\mathrm{Eve}}$, i.e. $\Psi _p^{\mathrm{Eve}} = \sum\nolimits_k {\rm{S}}_{p, k}^{\mathrm{Eve}} +\sum\nolimits_k  {\rm{I}}_{p, k}^{\mathrm{Eve}}$.
Next, due to convexity of the $-\phi_3$ term, the \ac{UL} sum secrecy rate is not necessarily concave. Thus, we apply the following lemma.\\\\
\textit{Lemma 3:} The term $\phi_3$ can be approximated to an affine function by applying Taylor expansion around $(\widetilde{\pmb{p}}, \{\widetilde{\pmb{V}}_l\}_{l=1}^L, \widetilde{\pmb{W}})$
\useshortskip
\begin{equation}
 \label{eq:UL-phi3}
	\begin{aligned}
		 \phi_3 & \approx \frac{1}{\ln 2}\sum_{p}\Big\{\ln\vert \widetilde{\Psi} _p^{\mathrm{Eve}} + \sigma_p^2\vert \\
        &+ \dfrac{\sum_k\vert g_{k,p} \vert^2{\Delta}_{p_k} +  \vert \alpha_p \vert^2\operatorname{tr}\left(\pmb{A}(\theta_p)\left[\sum_s\pmb{\Delta}_{\pmb{V}_s} + \pmb{\Delta}_{\pmb{W}}\right]\right)}{\widetilde{\Psi} _p^{\mathrm{Eve}} + \sigma_p^2}\Big\}.
	\end{aligned}
\end{equation}
where $\widetilde{\Psi} _p^{\mathrm{Eve}} = \sum\nolimits_k \widetilde{\rm{S}}_{p, k}^{\mathrm{Eve}} +\sum\nolimits_k  \widetilde{\rm{I}}_{p, k}^{\mathrm{Eve}}$ and
\useshortskip
\begin{align}
	\widetilde{\rm{S}}_{p, k}^{\mathrm{Eve}} &= \widetilde{p}_k \vert g_{k,p} \vert^2, \\
	\widetilde{\rm{I}}_{p, k}^{\mathrm{Eve}} &= \sum_{\substack{k'\neq k}}^K \widetilde{p}_{k'} \vert g_{k',p} \vert^2 + \vert \alpha_p \vert^2 \pmb{a}_{N_T}^H(\theta_p)\widetilde{\pmb{Q}}\pmb{a}_{N_T}(\theta_p)
\end{align}
\useshortskip
and $\widetilde{\pmb{Q}} = \sum_\ell \widetilde{\pmb{V}}_\ell + \widetilde{\pmb{W}}$.\\
\textbf{Proof:} The proof follows similar steps as that of \textit{Lemma 1}.

Therefore, the approximated lower bound concave \ac{UL} sum secrecy rate can be written as
\useshortskip
\vspace{-0.2cm}
\begin{equation}
\label{eq:LB-UL-sum-secrecy-rate}
\begin{aligned}
	&\wideparen{\SR}_{\mathrm{UL}}^{\mathrm{LB}}
	=\sum\limits_{k} \log_2(1+\varpi_k) +\sum\limits_{p,k} \log_2({\rm{I}}_{p, k}^{\mathrm{Eve}} + \sigma_p^2) \\
    &- \frac{K}{\ln 2}\sum_{p}\Big\{\ln\vert \widetilde{\Psi} _p^{\mathrm{Eve}} + \sigma_p^2\vert \\
    &+ \dfrac{\sum_k\vert g_{k,p} \vert^2{\Delta}_{p_k} +  \vert \alpha_p \vert^2\operatorname{tr}\left(\pmb{A}(\theta_p)\left[\sum_s\pmb{\Delta}_{\pmb{V}_s} + \pmb{\Delta}_{\pmb{W}}\right]\right)}{\widetilde{\Psi} _p^{\mathrm{Eve}} + \sigma_p^2}\Big\}.
\end{aligned}
\end{equation}
Now that the lower-bound of the \ac{UL} sum secrecy rate, i.e. $\wideparen{\SR}_{\mathrm{UL}}^{\mathrm{LB}}$, is concave, we can now formulate an efficient convex optimization problem that can be solved in the neighborhood of $(\widetilde{\pmb{p}}, \{\widetilde{\pmb{V}}_l\}_{l=1}^L, \widetilde{\pmb{W}})$. Indeed, combining the constraints in \eqref{eq:set-of-constraints-for-bounding-the-UL-sum-secrecy-rate} with problem $(\mathcal{P}_{1.2})$ in \eqref{eq:problem12}, and utilizing the concave \ac{DL} and \ac{UL} sum secrecy rates in \eqref{eq:SR_DL_bottom} and \eqref{eq:LB-UL-sum-secrecy-rate}, we arrive at the following convex optimization problem
									  

\useshortskip
\begin{equation}
 \label{eq:problem13}
\begin{aligned}
(\mathcal{P}_{1.3}):
\begin{cases}
\max\limits_{\lbrace \substack{\{\pmb{V}_\ell\}_{\ell}, \pmb{W}, \pmb{p},
                                     \{\varpi _k\}_k, \{\vartheta _k\}_k}  \rbrace} \wideparen{\SR}_{\mathrm{DL}} + \wideparen{\SR}_{\mathrm{UL}}^{\rm{LB}}&\\
\textrm{s.t.} \hspace{0.95cm} \sum_{\ell=1}^L\trace(\pmb{V}_{\ell}) \leq P_{\rm{DL}}^V,\ \trace(\pmb{W}) \leq P_{\rm{DL}}^W,&\\
  \quad \hspace{1.1cm}\sum\nolimits_{k=1}^K p_k \leq  P_{\rm{UL}},& \\

 \quad \hspace{1.1cm}\frac{\operatorname{tr}\left(\left(\sum_{\ell=1}^L \boldsymbol{V}_{\ell}+\boldsymbol{W}\right) \boldsymbol{A}_s\right)}{\operatorname{tr}\left(\left(\sum_{\ell=1}^L \boldsymbol{V}_{\ell}+\boldsymbol{W}\right) \boldsymbol{A}_m\right)} \leq \operatorname{ISMR}_{\max }   , &\\
\quad \hspace{1.1cm}\frac{\vartheta_k^2}{p_k} \leq \pmb{h}_{t,k}^H  {\boldsymbol{\Phi}}_k^{-1}\pmb{h}_{t,k} &\\
\quad \hspace{1.5cm}- \pmb{h}_{t,k}^H\widetilde{\boldsymbol{\Phi}}_k^{-1}\left(\boldsymbol{\Phi}_k-\widetilde{\boldsymbol{\Phi}}_k\right)\widetilde{\boldsymbol{\Phi}}_k^{-1} \pmb{h}_{t,k}\quad \forall k, &\\
\quad \hspace{1.1cm}\varpi_k \leq \widetilde{\vartheta}_k^2+2 \widetilde{\vartheta}_k\left(\vartheta_k-\widetilde{\vartheta}_k\right), \quad \forall k,&\\
\quad  \hspace{1.1cm}\pmb{W} \succeq \pmb{0}, \ \pmb{p} \succeq \pmb{0}, \pmb{V}_\ell \succeq \pmb{0}, \quad \forall \ell,&\\
\quad \hspace{1.1cm}\varpi _k \geq 0, \vartheta _k \geq 0, \quad \forall k.&
\end{cases}
\end{aligned}
\end{equation}
As the problem solves in the neighborhood of $(\widetilde{\pmb{p}}, \{\widetilde{\pmb{V}}_l\}_{l=1}^L, \widetilde{\pmb{W}})$, we should be able to quantify a good neighborhood to solve the problem at. For this, we design \ac{IJTB}, which leverages a \ac{BCCD} type method applied directly on $(\mathcal{P}_{1.3})$ given a previous solution of $(\widetilde{\pmb{p}}, \{\widetilde{\pmb{V}}_l\}_{l=1}^L, \widetilde{\pmb{W}})$, as well as the \ac{UL} beamforming vectors, and iterates between both the set of parameters $(\widetilde{\pmb{p}}, \{\widetilde{\pmb{V}}_l\}_{l=1}^L, \widetilde{\pmb{W}})$ and $\pmb{U}$. 
The \ac{IJTB} method is summarized in \textbf{Algorithm \ref{alg:alg1}}.

									  

%% file: sections/algorithm.tex
\begin{algorithm}[H]
\caption{IJTB for FD-ISAC sum secrecy UL/DL rate maximization}\label{alg:alg1}
\begin{algorithmic}
\STATE 
\STATE {\textbf{Input}: $\lbrace \pmb{h}_{r,\ell}, \sigma_\ell \rbrace_{\ell = 1}^L$, $\lbrace \pmb{h}_{t,k}  \rbrace_{k=1}^K$, $ \lbrace \gamma_p,\theta_p , \sigma_p, \alpha_p \rbrace_{p=1}^P$, $\lbrace g_{k,p} \rbrace_{p,k=1}^{P,K}$, $\ISMR_{\mathrm{max}}$, $\beta$, $\pmb{H}_{\mathrm{SI}}$, $\pmb{A}_s$, $\pmb{A}_m$. } 
\STATE \textbf{Initialize}:
\STATE \hspace{0.5cm}  Set $m=0$,  $\pmb{V}_\ell^{(0)}= \pmb{0}$, $\pmb{W}^{(0)} = \pmb{0}$, $\pmb{p}^{(0)} =  \pmb{0}$.
\STATE \textbf{while} $ m < M_{iter}$
\STATE \hspace{0.5cm} Update $\widetilde{\pmb{Q}}$ as $\widetilde{\pmb{Q}} =\sum\nolimits_{\ell=1}^L \widetilde{\pmb{V}_\ell} + \widetilde{\pmb{W}}$.
\STATE \hspace{0.5cm} \textbf{for} $k = 1 \ldots K$
\STATE \hspace{1.0cm} Update $\pmb{\Phi}_k^{(m)}$ via \eqref{eq:phi_k_eqn} given $\widetilde{\pmb{Q}}$ and $\widetilde{\pmb{p}}$.
\STATE \hspace{1.0cm} Update $\pmb{u}_k^{(m+1)}$ as $\pmb{u}_k^{(m+1)} 
	=
	\left(\pmb{\Phi}_k^{(m)}\right)^{-1} 
	 \pmb{h}_{t,k}$.
\STATE \hspace{1.0cm} Normalize $\pmb{u}_k^{(m+1)}$ as $\pmb{u}_k^{(m+1)} \leftarrow \pmb{u}_k^{(m+1)} / \left|\pmb{u}_k^{(m+1)}\right|
$.
  
\STATE \hspace{0.5cm} Solve $(\mathcal{P}_{1.3}^{(m)})$ in \eqref{eq:problem12} to get $\pmb{V}_\ell^{(m+1)},\pmb{W}^{(m+1)}, \pmb{p}^{(m+1)}$.
  
\STATE \textbf{set} $\widetilde{\pmb{W}} = \pmb{W}^{(m)}$, $\widetilde{\pmb{p}} = \pmb{p}^{(m)}$, $\lbrace \widetilde{\pmb{V}}_\ell = {\pmb{V}}_\ell^{(m)} \rbrace_{\ell = 1}^L$, $\lbrace \widetilde{\vartheta}_k = {\vartheta}_k^{(m)} \rbrace_{k = 1}^K$.
\STATE \textbf{return}  $\lbrace \pmb{u}_k  \rbrace_{k=1}^K$, $\pmb{W}$, $\pmb{p}$, $\pmb{V}$.
\end{algorithmic}
\end{algorithm}

%% file: sections/simulation-results.tex
In this section, we present our simulation results. 
Before we discuss and analyze and our findings, we mention the simulation parameters and benchmarks used in simulations.
\subsection{Parameter Setup}
\input{sections/simulation-parameters.tex}
\input{sections/simulation-parameters-table.tex}

\subsection{Benchmark Schemes}
Throughout simulations, we compare our the \ac{IJTB} method with the following schemes:
\begin{itemize}
	\item 
\textbf{Benchmark 1}: In this benchmark, partial transmit power is allocated to emit the \ac{AN} to interfere with the eavesdroppers. In fact, the \ac{AN} is isotropically distributed on the orthogonal complement subspace of the \ac{DL} users \cite{1003835}
This benchmark is denoted by \textbf{Iso AN}. In other words, the power is distributed equally between the \ac{AN}, the \ac{DL} beamforming and the uplink users. 
	\item 
\textbf{Benchmark 2}: This benchmark assumes on \ac{AN} and thus the total power budget is distributed equally amongst the \ac{DL} beamforming and the \ac{UL} users.
This benchmark is denoted by \textbf{Iso No-AN}.

\item 
\textbf{Benchmark 3}: This benchmark is denoted by \textbf{Feasible}
 This is a random baseline scheme utilizing all the available power budgets available with equality.
 More precisely, the beamforming vectors, in conjunction with the power vectors and artificial noise statistics, are randomly generated. 
 Such a scheme is included to show how a random and simple scheme can perform in a secure \ac{FD}-\ac{ISAC} system.
 We emphasize on term feasible, as it is a random solution satisfying the optimization problem.
 Random benchmark was adopted in several works like \cite{10494408,10041957}.
\end{itemize} 

\subsection{Simulation Results}
\begin{figure}[t]
	\centering
	\includegraphics[width=1\linewidth]{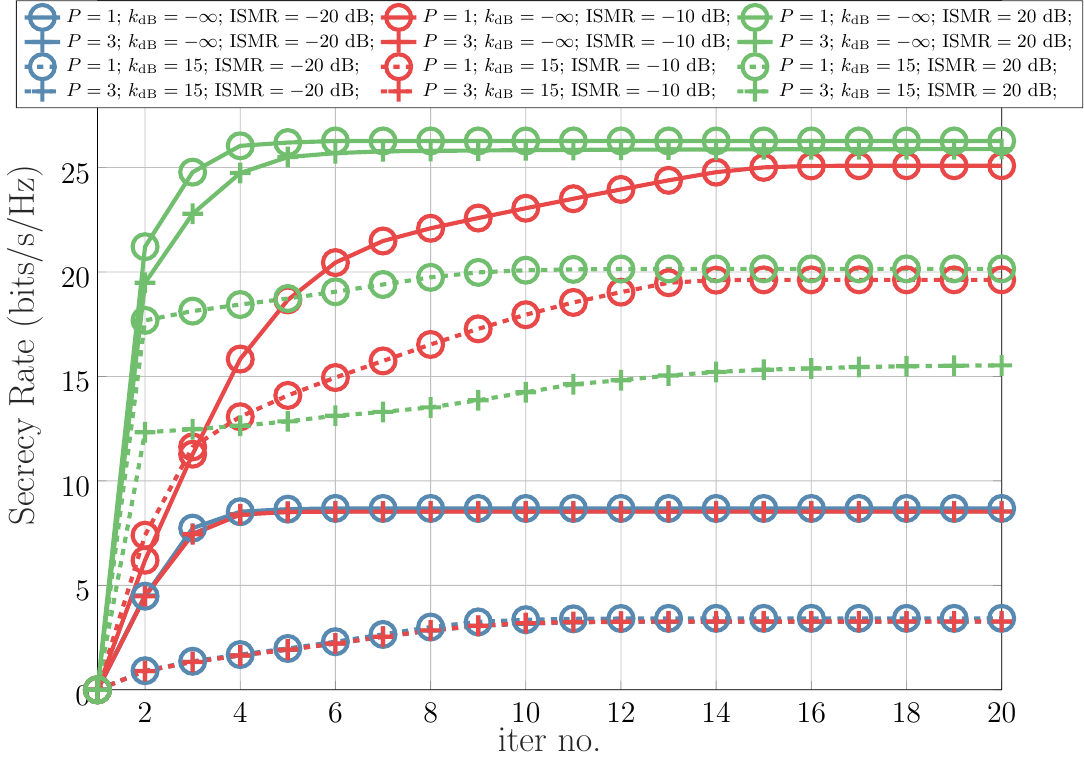}
	\caption{The convergence behavior of the \ac{IJTB} algorithm for different $P$ and $\ISMR_{\rm{max}}$ levels (denoted as $\ISMR$ for short).
We set $N_T = 12$, 
$N_R = 6$, 
$K = 2$, 
$L = 1$, 
$\sigma_n^2 = \sigma_\ell^2 =-70 \dB$,
$\sigma_p^2 = -65 \dB$,
$P_{\rm{DL}}^V = 0\dB$, 
$r_{\rm{DL}} = 20 \meters$,
$r_{\rm{UL}} = 15 \meters$, and
$r_{\rm{Eve}} = 17 \meters$.}
	\label{fig:iter01}
\end{figure}
In Fig. \ref{fig:iter01}, we set 
$N_T = 12$ antennas, 
$N_R = 6$ antennas, 
$K = 2$ users, 
$L = 1$ users, 
$\sigma_n^2 = -70 \dB$,
$\sigma_\ell^2 = -70 \dB$,
$\sigma_p^2 = -65 \dB$,
$P_{\rm{DL}}^V = 0\dB$, 
$r_{\rm{DL}} = 20 \meters$,
$r_{\rm{UL}} = 15 \meters$, and
$r_{\rm{Eve}} = 17 \meters$.
We plot the sum secrecy rate as a function of iteration number, with the intent of studying convergence with different channel conditions, $\ISMR_{\rm{max}}$ levels, and number of target eavesdroppers involved in the \ac{FD}-\ac{ISAC} setup. 
On the other hand, generally speaking, at \ac{NLoS} conditions, the sum secrecy rate is better than the \ac{LoS} ones. 
For instance, at \ac{NLoS} conditions ($k_{\rm{dB}} = -\infty$), when $P = 3$ and $\ISMR = 20\dB$, the sum secrecy rate converges to $25.88 \bpspHz$ whereas at strong \ac{LoS} ($k_{\rm{dB}} = 15 \dB$), we see that for the same conditions, the sum secrecy rate settles at about $15.52 \bpspHz$.
Also, when we configure a less stringent \ac{ISMR} constraint onto the proposed \ac{IJTB} method, we notice that the sum secrecy rate improves and takes lesser iterations to convergence to its maximum value.
For example, at \ac{NLoS} and for $P = 1$ eavesdroppers and at $\ISMR = 20 \dB$, the algorithm converges to an overall sum secrecy rate of about $ 26.27 \bpspHz$, and by lowering the \ac{ISMR} to $-10\dB$, the sum secrecy rate slightly decreases to $25.08 \bpspHz$. Now, when \ac{ISMR} is set to a stringent level of $-20\dB$, we observe that the sum secrecy rate dramatically drops to $8.67\bpspHz$.
This is because more emphasis is being imposed on the sensing counterpart of the \ac{ISAC} system, hence more sensing priority can impact the sum secrecy rate performance.
It is also worth noting that at \ac{NLoS} conditions and at a less stringer \ac{ISMR}, the sum secrecy rate of both $P = 1$ and $P = 3$ are almost the same.
We observe that more eavesdroppers impact the sum secrecy rate in a negative way. In particular, more eavesdroppers lower the sum secrecy rate at convergence. 
For instance, at $k_{\rm{dB}} = 15 \dB$, and $\ISMR = 20\dB$, with $P = 1$ the sum secrecy rate converges to $20.15\bpspHz$ whereas for $P = 3$, the sum secrecy rate drops to $15.52\bpspHz$.
This is because of a fixed power budget overall the scenarios.
In any case, we can observe algorithmic convergence in about $10-14$ iterations at most.

\begin{figure}[t]
	\centering
	\includegraphics[width=1\linewidth]{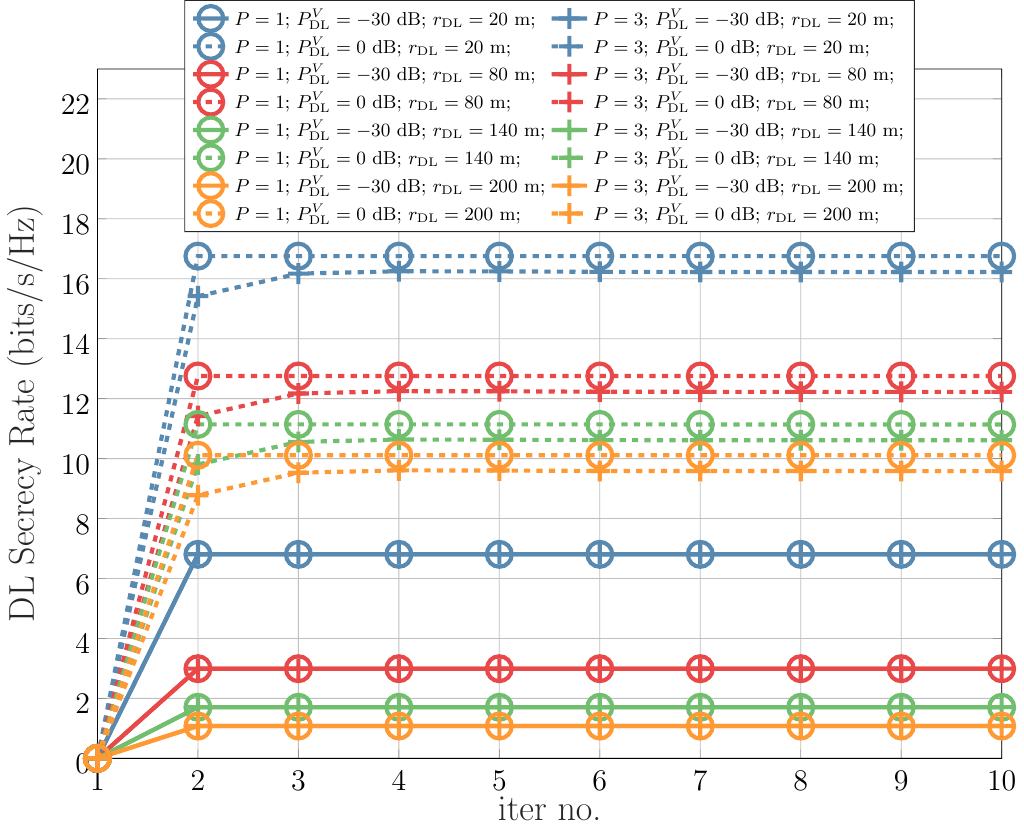}
	\caption{\ac{DL} sum secrecy rate convergence of the \ac{IJTB} method for different $P$, power budget $P_{\rm{DL}}^V$ and \ac{DL} user distances, $r_{\rm{DL}}$. We set the same parameters as Fig. \ref{fig:iter01} except $k_{\rm{dB}} = 15 \dB$, $\ISMR_{\rm{max}} = 20 \dB$, and $r_{\rm{Eve}} = 175 \meters$.}
	\label{fig:iter03}
\end{figure}
In Fig. \ref{fig:iter03}, we set the same parameters as those of Fig. \ref{fig:iter01} with the exception of 
$k_{\rm{dB}} = 15 \dB$,
$\ISMR_{\rm{max}} = 20 \dB$, and 
$r_{\rm{Eve}} = 175 \meters$.
We study the impact of \ac{DL} power budget and $r_{\rm{DL}}$ on the \ac{DL} sum secrecy rate as a function of iteration number.
We notice that as the distance between the \ac{DL} user and the \ac{DFRC} \ac{BS} decreases, the algorithm tends to converge to a higher \ac{DL} sum secrecy rate.
For example, when $P = 1$, and only $P_{\rm{DL}}^V = -30 \dB$ is utilized for power budget, the \ac{IJTB} algorithm converges to $1.07 \bpspHz$ at a high distance of $r_{\rm{DL}} = 200 \meters$.
Meanwhile, when the same distance is lowered to $140 \meters$, the \ac{DL} sum secrecy rate tends to $1.7 \bpspHz$. 
Moreover, this sum secrecy rate increases to $3 \bpspHz$ at $r_{\rm{DL}} = 80 \meters$, and $6.8 \bpspHz$ at $r_{\rm{DL}} = \meters $.
This is because of the increases pathloss effect at higher distances at a limited power budget.
Moreover, the impact of $P_{\rm{DL}}^V$ also translates to a boost in \ac{DL} sum secrecy rate. Generally speaking, we can see that an increase of $30\dB$ in \ac{DL} power budget translates to a \ac{DL} sum secrecy rate improvement of roughly $10 \bpspHz$. 
This is because allocating more power within the system at a constant level of $\ISMR_{\rm{max}}$ level can only improve the secrecy.
In addition, we see that at fixed power budget and $\ISMR_{\rm{max}}$ level, more target eavesdroppers in the system deteriorates the \ac{DL} sum secrecy rate. 
At $P_{\rm{DL}}^V = 0 \dB$, we observe that tripling the number of eavesdroppers leads to a decreases of \ac{DL} sum secrecy rate of roughly $0.52 \bpspHz$.
However, for a low power budget (i.e. $P_{\rm{DL}}^V = -30 \dB$), we notice that the algorithm converges to the same \ac{DL} sum secrecy rate for both $P=1$ and $P=3$ cases. 
This is because the \ac{ISAC} system is at higher risk of potential eavesdropping from a subset of eavesdroppers, who experience better channel conditions.
In this case, we see that the algorithm needs no more than $3$ iterations to converge.

\begin{figure}[t]
	\centering
	\includegraphics[width=1\linewidth]{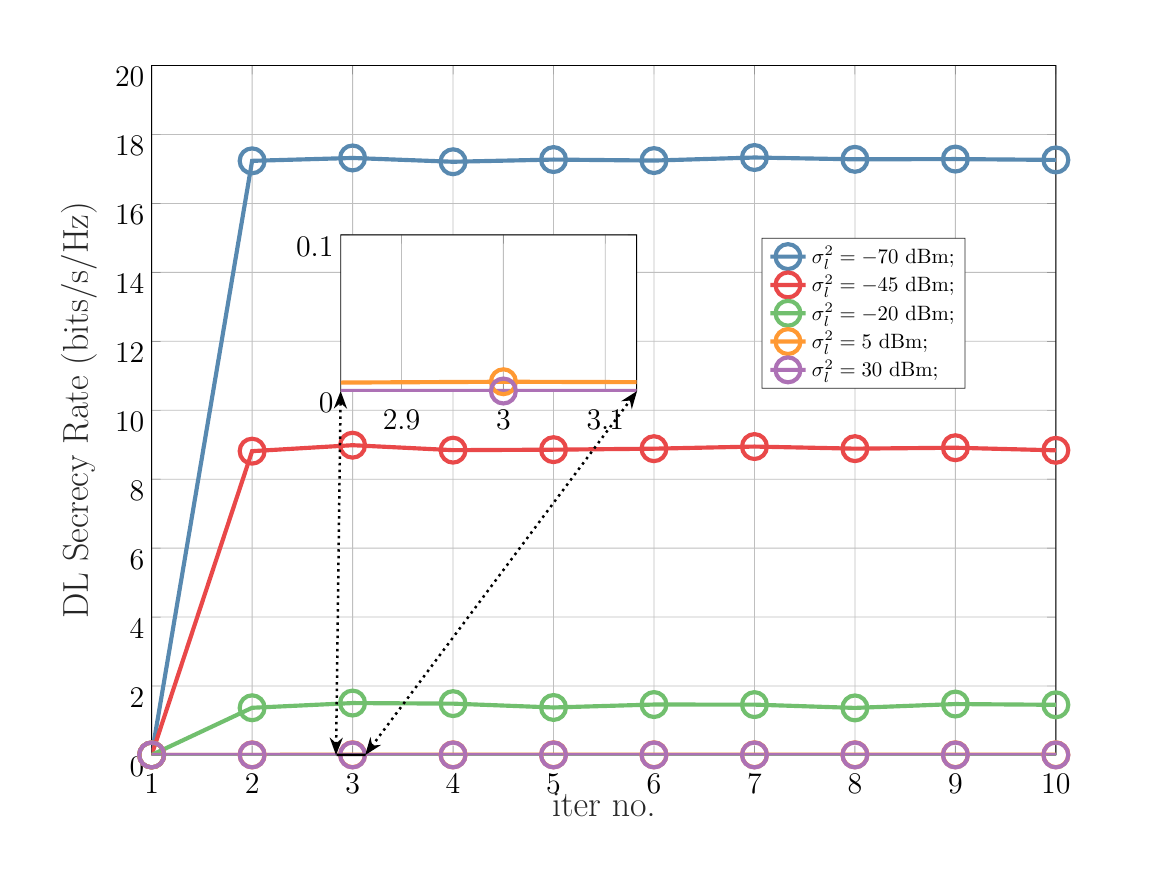}
	\caption{The \ac{DL} sum secrecy rate convergence behavior of the \ac{IJTB} algorithm for different noise \ac{DL} noise variances, i.e. $\sigma_l^2$. We fix the same parameters as Fig. \ref{fig:iter01} except for $P = 1$, $k_{\rm{dB}} = -\infty \dB$, and $\ISMR_{\rm{max}} = 20 \dB$.}
	\label{fig:iter05}
\end{figure}
In Fig. \ref{fig:iter05}, we set the same simulation parameters as Fig. \ref{fig:iter01} except for $P = 1$, $k_{\rm{dB}} = -\infty \dB$, and $\ISMR_{\rm{max}} = 20 \dB$. We study the impact of $\sigma_l^2$ on the \ac{DL} sum secrecy rate with increasing iteration number.
The more noisy a \ac{DL} user, the less of a \ac{DL} sum secrecy rate the algorithm settles at. 
For instance, for a $\sigma_l^2 = -70 \dBm$, the \ac{DL} sum secrecy rate converges to $17.27 \bpspHz$ as opposed to $8.88 \bpspHz$ at $\sigma_l^2 = -45 \dBm$. When the noise variance exceeds $5\dBm$, which is already too high, the \ac{DL} sum secrecy rate is near-zero.
This is because at a fixed power budget and \ac{ISMR} level, the \ac{DL} noise variance can only decrease the \ac{DL} sum secrecy rate, which is an intuitive observation.
We also observe, as in Fig. \ref{fig:iter01} and Fig. \ref{fig:iter03}, that the algorithm converges in all cases. 

\begin{figure}[t]
	\centering
	\includegraphics[width=1\linewidth]{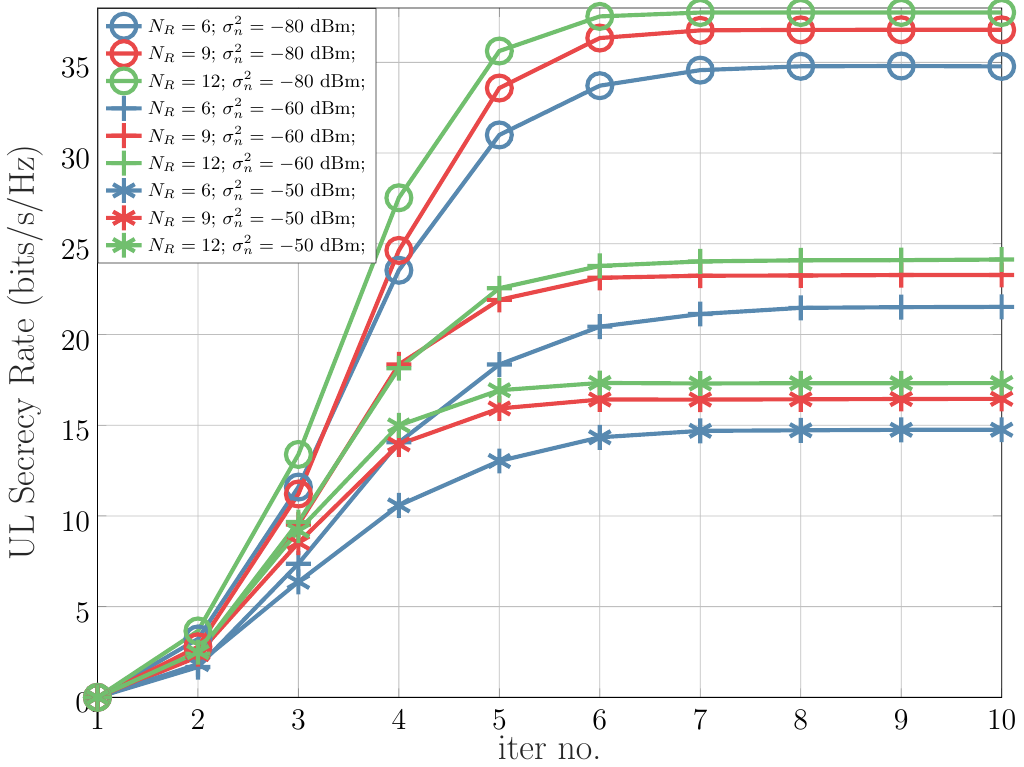}
	\caption{\ac{UL} sum secrecy rate convergence of the \ac{IJTB} method for different number of receive antennas $N_R$ and noise variance at the \ac{DFRC} \ac{BS}. We fix the same parameters as Fig. \ref{fig:iter05} except for $N_T = 6$ and $\sigma_\ell^2 = -100 \dB$.}
	\label{fig:iter04}
\end{figure}
In Fig. \ref{fig:iter04}, we set the same simulation parameters as Fig. \ref{fig:iter05} except for $N_T = 6$ antennas and $\sigma_\ell^2 = -100 \dB$.
We study the impact of both number of \ac{DFRC} \ac{BS} receive antennas, i.e. $N_R$, and the noise variance at \ac{DFRC} \ac{BS}, i.e. $\sigma_n^2$, on the convergence behavior of the \ac{UL} sum secrecy rate.
For any $\sigma_n^2$, we see that increasing the number of receive antennas can improve the \ac{DL} sum secrecy rate for fixed power budget, and fixed \ac{ISMR} levels. 
For instance, at fixed $\sigma_n^2 = -50 \dBm$, we see that with $N_R = 6$ antennas, the \ac{IJTB} algorithm converges to an \ac{UL} sum secrecy rate of about $14.71 \bpspHz$, as compared to $16.44 \bpspHz$ for $N_R = 9$ antennas, and $17.32 \bpspHz$ for $N_R = 12$ antennas.
This is because \ac{DFRC} receive antennas can also allow room for enhanced degrees-of-freedom which enables better \ac{UL} sum secrecy rate for the algorithm to converge towards.
We observe that the algorithm needs about $8$ iterations to converge to a stable point.

\begin{figure}[t]
	\centering
	\includegraphics[width=1\linewidth]{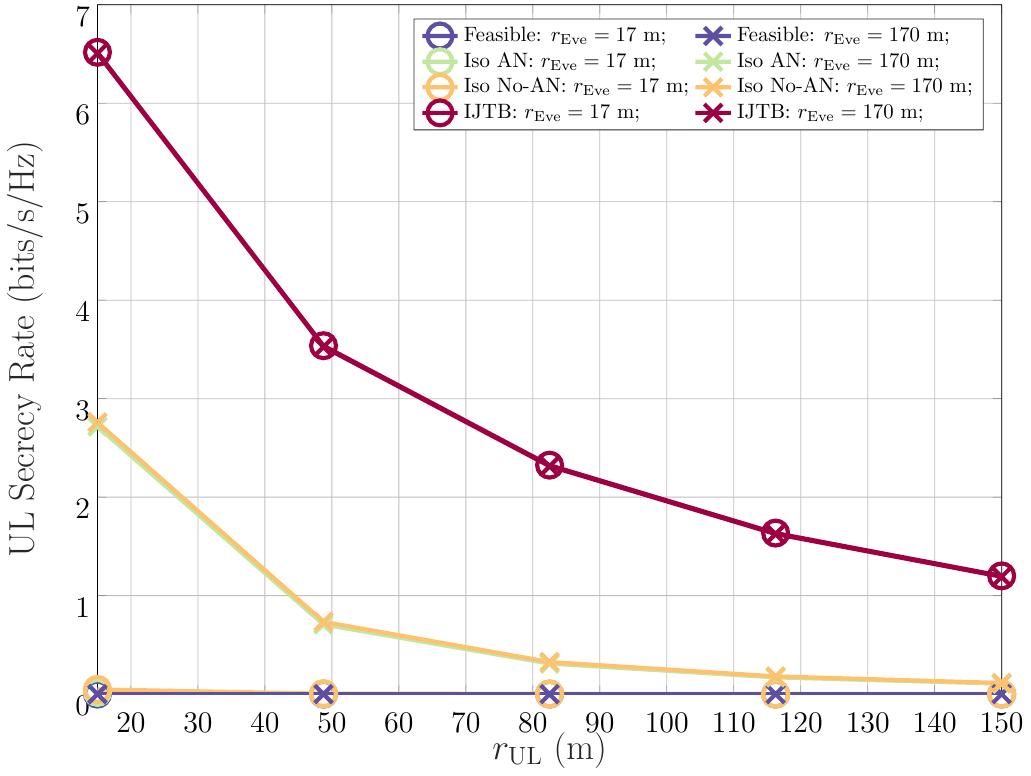}
	\caption{\ac{UL} sum secrecy rate performance for different target eavesdropper distances, i.e. $r_{\rm{Eve}}$, as a function of \ac{UL} user distances, i.e. $r_{\rm{UL}}$. We fix the same  parameters as Fig. \ref{fig:iter01} except for $N_R = 12$,
$K = 1$,
$P = 1$,
$k_{\rm{dB}} = -\infty \dB$,
$\sigma_\ell^2 = -100 \dB$, and
$\ISMR_{\rm{max}} = 20 \dB$.}
	\label{fig:noniter01}
\end{figure}
In Fig. \ref{fig:noniter01}, we set the same simulation parameters as Fig. \ref{fig:iter01} except for the following: 
$N_R = 12$ antennas,
$K = 1$ user,
$P = 1$ evesdroppers,
$k_{\rm{dB}} = -\infty \dB$,
$\sigma_\ell^2 = -100 \dB$, and
$\ISMR_{\rm{max}} = 20 \dB$.
We study the performance of the \ac{IJTB} method from an \ac{UL} sum secrecy rate sense with increasing \ac{UL} radial distance and for different eavesdropper distances.
It is evident that with increasing $r_{\rm{UL}}$, the \ac{UL} decreases regardless of the method.
However, it is worthwhile noting that the \ac{IJTB} method dominates all other benchmarks regardless of $r_{\rm{UL}}$ and $r_{\rm{Eve}}$. 
Even more, the \ac{IJTB} method can adapt to different $r_{\rm{Eve}}$, which is not a common feature with the other benchmark. For instance, the \ac{UL} sum secrecy rate of all benchmarks drops to a low near-zero rate when the eavesdropper reaches a distance of $r_{\rm{Eve}} = 17 \meters$.

\begin{figure}[t]
	\centering
	\includegraphics[width=1\linewidth]{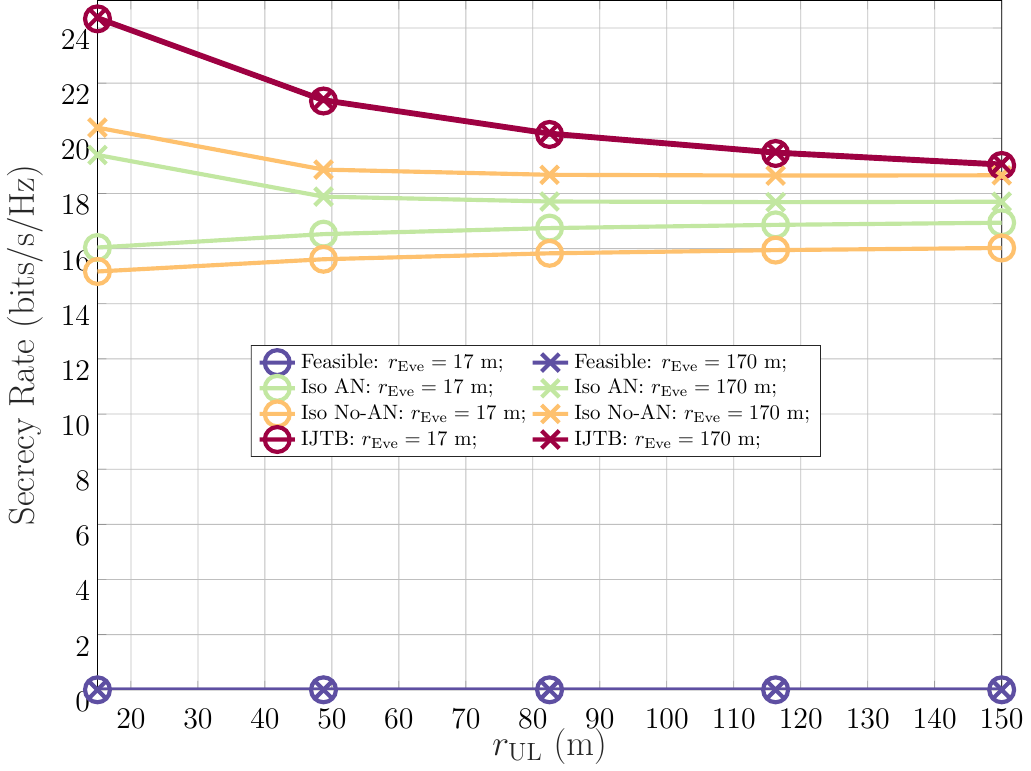}
	\caption{Overall sum secrecy rate performance for different target eavesdropper distances, i.e. $r_{\rm{Eve}}$, and for various \ac{UL} user distances. We fix the same simulation parameters as Fig. \ref{fig:noniter01}.}
	\label{fig:noniter02}
\end{figure}
In Fig. \ref{fig:noniter02}, we fix the same simulation parameters as that of Fig. \ref{fig:noniter01}. 
Furthermore, we analyze the efficacy of the \ac{IJTB} approach in terms of overall sum secrecy rate as the \ac{UL} user distance increases, considering various distances of potential eavesdroppers.
We observe that the sum secrecy rate decreases with \ac{UL} sum secrecy rate for all methods. In addition, a far eavesdropper would normally contribute to an increase in sum secrecy rate.
This is because of increased pathloss. 
What's interesting to observe is that for a close eavesdropper (i.e. $r_{\rm{Eve}} = 17 \meters$), the impact of \ac{AN} is more effective, which is seen by the two isotropic benchmarks (isotropic and non-isotropic benchmarks), where there is about $0.84 \bpspHz$ gap in favor of the isotropic \ac{AN} benchmark. 
However, when the eavesdropper distance becomes high (i.e. $r_{\rm{Eve}} = 170 \meters$), the impact of \ac{AN} is no longer effective. 
It is also interesting to observe that any random solution of power vector, beamformers and \ac{AN} beamforming on average keeps the sum secrecy rate at zero.
On the other hand, the \ac{IJTB} method can adapt to eavesdropper distance changes as can be seen because both curves corresponding to the \ac{IJTB} method at the two different eavesdropper distances coincide. 
Even more, the \ac{IJTB} method outperforms all other methods in terms of sum secrecy rate over all the evaluated \ac{UL} user distances.
For instance, when $r_{\rm{UL}} = 50 \meters$, the \ac{IJTB} method achieves a sum secrecy rate of about $21.4 \bpspHz$ as opposed $18.86 \bpspHz$ achieved by the isotropic no-\ac{AN} benchmark.

\begin{figure}[t]
	\centering
	\includegraphics[width=1\linewidth]{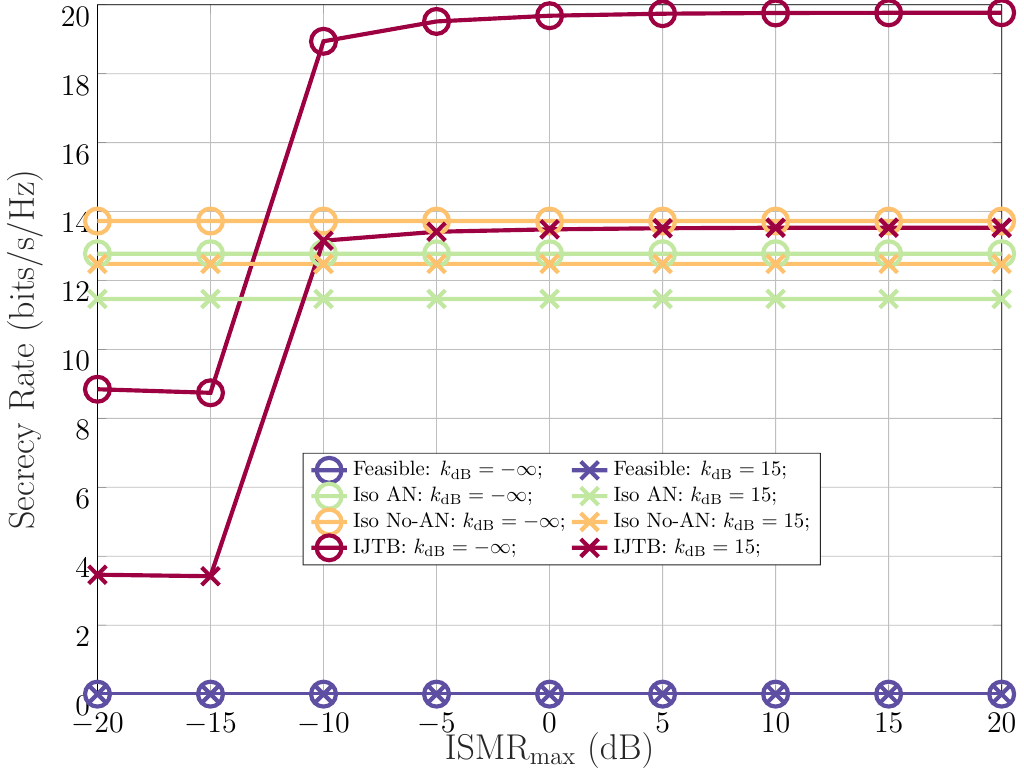}
	\caption{The sum secrecy rate-\ac{ISMR} trade-off for Rayleigh and strong \ac{LoS} channels. We set the same parameters as Fig. \ref{fig:iter01} except for $P = 1$, $r_{\rm{DL}} = 200 \meters$, and $r_{\rm{Eve}} = 175 \meters$.}
	\label{fig:noniter03}
\end{figure}
In Fig. \ref{fig:noniter03}, we fix the same simulation parameters as Fig. \ref{fig:iter01} except for
$P = 1$ eavesdropper,
$r_{\rm{DL}} = 200 \meters$, and
$r_{\rm{Eve}} = 175 \meters$.
We study the sum secrecy rate vs. \ac{ISMR} \ac{ISAC} trade-off for different channel conditions.
When the \ac{ISMR} constraint imposes a stringent restriction through a tight (low) $\rm{ISMR}_{\rm{max}}$ the sum secrecy rate of the \ac{IJTB} method drops lower than the isotropic benchmarking methods. 
This is because the \ac{IJTB} method is basically dedicating an optimized chunk of power towards the eavesdropper through the mainlobes, while restricting power onto the sidelobes.
However, it is interesting to see that the \ac{IJTB} method is still able to achieve high sum secrecy rates, which are, in some channel conditions, comparable with the benchmarks.
For example, in Rayleigh channel conditions ($k_{\rm{dB}} = -\infty$) and at a tight \ac{ISMR} corresponding to $\rm{ISMR}_{\rm{max}} = -20\dB$, the sum secrecy rate is about $8.84 \bpspHz$ and as the \ac{ISMR} constraint becomes less and less stringent (i.e. increasing $\rm{ISMR}_{\rm{max}}$), the sum secrecy rate of the \ac{IJTB} method gradually increases to increase the maximum possible sum secrecy rate the \ac{ISAC} system can achieve.
For example, the sum secrecy rate reaches $19.76 \bpspHz$ and $13.53 \bpspHz$ at high $\rm{ISMR}_{\rm{max}}$ for Rayleigh and strong \ac{LoS} channel conditions, respectively.  
 This is because the channel conditions hold significant amount of information regarding the sum secrecy rate of the \ac{IJTB} system.
Clearly, the sum secrecy rates of all other benchmarks do not depend on the $\rm{ISMR}_{\rm{max}}$ because they are not optimized for good \ac{ISMR} performance.

\begin{figure}[t]
	\centering
	\includegraphics[width=1\linewidth]{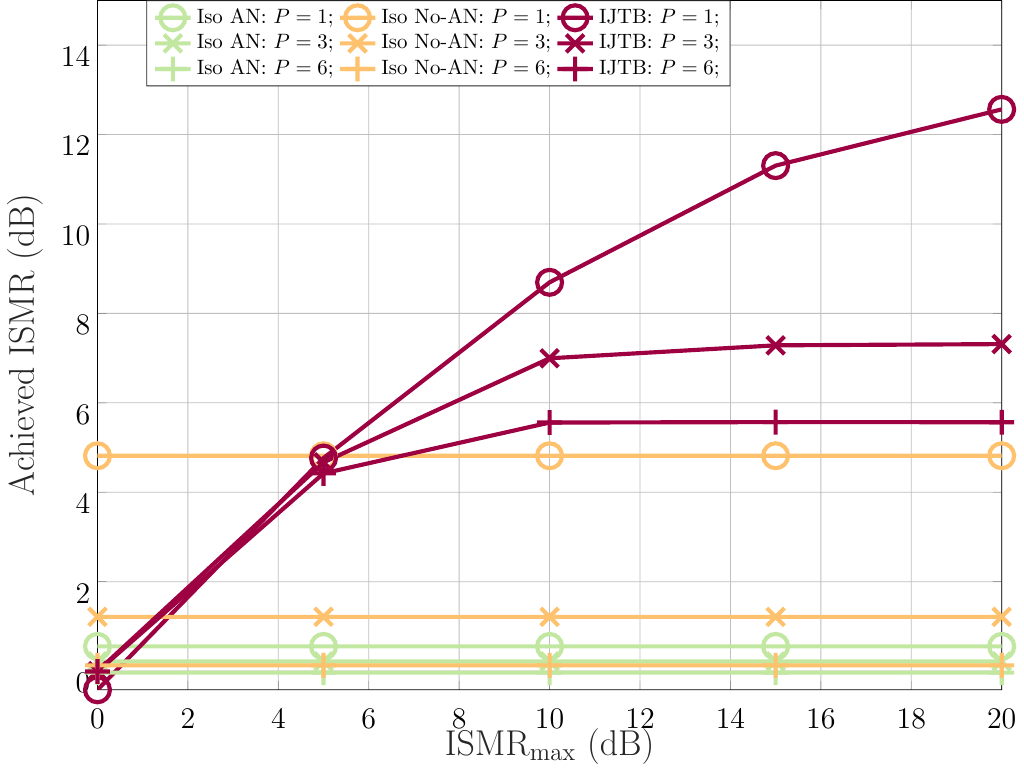}
	\caption{The achieved \ac{ISMR} as a function of $\ISMR_{\rm{max}}$ for different number of eavesdroppers, $P$. We set the same parameters as Fig. \ref{fig:iter01} except for $k_{\rm{dB}} = -\infty \dB$ and $r_{\rm{Eve}} = 175 \meters$.}
	\label{fig:noniter04}
\end{figure}
In Fig. \ref{fig:noniter04}, we set the same simulation parameters as Fig. \ref{fig:iter01} except for
$k_{\rm{dB}} = -\infty \dB$ and
$r_{\rm{Eve}} = 175 \meters$.
Moreover, we analyze the impact of the $\ISMR_{\rm{max}}$ on the achieved \ac{ISMR} of the \ac{IJTB} method. 
We can see a clear trend between the obtained \ac{ISMR} and the tuned one via $\ISMR_{\rm{max}}$.
Notice that the evaluated \ac{ISMR} is always upper bounded by the given $\ISMR_{\rm{max}}$, which tells us that this constraint is always active.
Interestingly, as $P$ increases, the \ac{ISMR} tries to hold on to a tighter value.
For example, for $P = 1$ eavesdroppers, when we set an $\ISMR_{\rm{max}}$ of $10 \dB$, the achieved \ac{ISMR} is $8.69 \dB$, as opposed to $7\dB$ for $P = 3$ eavesdroppers and $5.55 \dB$ for $P = 6$ eavesdroppers. Also, when $\ISMR_{\rm{max}} > 10 \dB$, we see that the case of $P = 6$ eavesdroppers saturates at about $5.55 \dB$.
This is because more energy has to be contained within the mainlobes, as compared to the sidelobes, towards the different target eavesdroppers.


%% file: sections/simulation-parameters.tex
We have $1$ cell, with cell radius of $500 \meters$ \cite{8340227}.
The \ac{DFRC} \ac{BS} is situated at the center of the cell (origin) of the cell, with transmit power budget of $40\dBm$ \cite{8207426}.
The transmit/receive array geometries at the \ac{DFRC} \ac{BS} follow a \ac{ULA} structure with antenna spacing of $\frac{\lambda}{2}$, where $\lambda$ is wavelength corresponding to a carrier frequency of $28\GHz$. The transmit antenna gain power of the \ac{DFRC} is set to $25\dBi$ \cite{8207426}.
Moreover, the receive antenna gain of the \ac{DFRC} has a gain of $25\dBi$ \cite{8207426}.
On the other hand, the \ac{DL} users are equipped with a single-antenna of $12\dBi$ antenna gain \cite{8207426} and the \ac{UL} users are also single-antenna users with $17\dBi$ \cite{8207426} antenna gain.
For fairness, we treat the evesdropper meaning that they are also single-antenna users with an antenna gain of $12\dBi$.
The residual \ac{SI} is set to $-110\dB$, which can be achieved over \ac{RF} and digital domains of the transceiver \cite{6832464}, \cite{9887801}.
The power budget of the proposed optimization method in the \ac{UL} and the \ac{AN} \ac{DL} are set to $P_{\rm{UL}} = P_{\rm{DL}}^W = 0\dB$.
The thermal noise is set to $-174\dBmpHz$.
Also, we utilize Rician  \cite{8207426} and Rayleigh \cite{9217488} distributions in order to simulate various channel conditions, with log-normal shadowing of $20\dB$. 

%% file: sections/simulation-parameters-table.tex
\begin{table}[!t]
\caption{Simulation Parameters\label{tab:table1}}
\centering
{
\begin{tabular}{|c||c|}
\hline
\textbf{Parameter} & \textbf{Value}\\
\hline
Number of cells & $1$ \\
\hline
Cell radius & $500\meters$ \cite{8340227} \\
\hline
Transmit power & $40\dBm$ \cite{8207426}  \\
\hline
Residual \ac{SI} & $-110\dB$ \cite{6832464,9887801} \\
\hline
Antenna geometry & \ac{ULA} \\
\hline
Antenna spacing & $\frac{\lambda}{2}$ \\
\hline
Channel conditions & \makecell{Rayleigh ($k_{\rm{dB}} = -\infty\dB$) \cite{9217488},\\ Rician ($k_{\rm{dB}} = 15\dB$) \cite{8207426}} \\
\hline
DFRC transmit antenna gain & $25\dBi$ \cite{8207426}  \\
\hline
\ac{UL} user antenna gain & $17\dBi$ \cite{8207426}  \\
\hline
DFRC receive antenna gain & $25\dBi$ \cite{8207426}  \\
\hline
\ac{DL} receive antenna gain & $12\dBi$ \cite{8207426}  \\
\hline
Eavesdropper receive antenna gain & $12\dBi$ \cite{8207426}  \\
\hline
Carrier frequency & $28\GHz$  \\
\hline
Thermal noise& $-174\dBmpHz$ \\
\hline 
Shadowing & log-normal of $20\dB$ \\
\hline 
Pathloss exponent & $2$ (free space) \\
\hline 
no. of DFRC transmit antennas $N_T$ & $\lbrace 6,12 \rbrace$ antennas \\
\hline 
no. of DFRC receive antennas $N_R$ & $\lbrace 6,9,12 \rbrace$ antennas \\
\hline 

\end{tabular}
}
\end{table}

%% file: sections/conclusion.tex
This paper examines an FD-ISAC system with multiple malicious targets aiming to intercept UL and DL communications and provides an \ac{ISAC} optimization framework for optimizing the UL and DL sum secrecy rate under sensing and power budget constraints.
This enables one quantifying the maximum achievable FD secrecy rates, under ISAC constraints.
Furthermore, we introduced IJBT, a method well-suited for the proposed FD-ISAC sum secrecy rate maximization problem at hand, which alternates between sub-problems to efficiently solve for beamformers and power allocations. 
Our results support the effectiveness of the method, as compared to benchmarking techniques.
In conclusion, our findings demonstrate that the IJTB algorithm not only exhibits rapid convergence, but also consistently outperforms other methods by maintaining higher sum secrecy rates, even as the eavesdropper distance increases. Furthermore, IJTB maintains strong performance, particularly as the \ac{ISMR} constraint becomes less stringent, highlighting its robustness in securing \ac{SAC} applications.

%% file: sections/appendix1.tex
The Taylor series expansion of $\phi_1$ around $\mathscr{S} \triangleq \left( \widetilde{\pmb{p}}, \{\widetilde{\pmb{V}}_{\ell}\}_{\ell = 1}^L,\widetilde{\pmb{W}} \right)$ can be expressed as
\useshortskip
\begin{equation}
	\begin{aligned}
& \phi_1\left(\boldsymbol{p}, \{\boldsymbol{V}_{\ell}\}_{\ell = 1}^L, \boldsymbol{W}\right) \\ &= \phi_1\left(\widetilde{\boldsymbol{p}},\{ \widetilde{\boldsymbol{V}}_{\ell}\}_{\ell = 1}^L,\widetilde{\boldsymbol{W}}\right) + \sum_{\ell}\left\{\sum_k \left.\frac{\partial \phi_1}{\partial p_k}\right|_{\mathscr{S}} (p_k - \widetilde{p}_k)\right. \\
& +\sum_{s}\left[\left.\frac{\partial \phi_1}{\partial \operatorname{Re}\left[\boldsymbol{V}_{s}\right]}\right|_{\mathscr{S}},\left.\frac{\partial \phi_1}{\partial \operatorname{Im}\left[\boldsymbol{V}_{s}\right]}\right|_{\mathscr{S}}\right]\left[\begin{array}{c}
\operatorname{Re}\left(\boldsymbol{V}_{s}-\widetilde{\boldsymbol{V}}_{s}\right)\\
\operatorname{Im}\left(\boldsymbol{V}_{s}-\widetilde{\boldsymbol{V}}_{s} \right)
\end{array}\right]\\
& +\left[\left.\frac{\partial \phi_1}{\partial \operatorname{Re}[\boldsymbol{W}]}\right|_{\mathscr{S}},\left.\frac{\partial \phi_1}{\partial \operatorname{Im}[\boldsymbol{W}]}\right|_{\mathscr{S}}\right]\left[\begin{array}{c}
\operatorname{Re}\left(\boldsymbol{W}-\widetilde{\boldsymbol{W}}\right) \\
\operatorname{Im}\left(\boldsymbol{W}-\widetilde{\boldsymbol{W}}\right)
\end{array}\right]\Bigg\}.
\end{aligned}
\end{equation}
The partial derivatives are trivial to calculate since every term appearing in $\phi_1$ is linear. 
This can be easily rearranged as follows
\useshortskip
\begin{equation}
	\begin{aligned}
		& \phi_1\left(\boldsymbol{p}, \{\boldsymbol{V}_{\ell}\}_{\ell = 1}^L, \boldsymbol{W}\right) \\ 
  & = \sum_{\ell}\log _2\left|{\textrm{I}}_{\ell}^{\mathrm{DL}}\right| + \sum_{\ell} \frac{1}{\ln 2} \frac{1}{{\textrm{I}}_{\ell}^{\mathrm{DL}}} \left\{ \sum_k \vert q_{k,\ell} \vert^2\Delta_{p_k}\right.\\
	&+ \sum_{s \neq \ell} \left\{\operatorname{tr}\left[\pmb{H}_{r,\ell}\operatorname{Re}\left(\Delta _{\pmb{V}_s}\right) + j\pmb{H}_{r,\ell}\operatorname{Im}\left(\Delta _{\pmb{V}_s}\right)\right]\right\} \\
	&+ \operatorname{tr}\left[\pmb{H}_{r,\ell}\operatorname{Re}\left(\Delta _{\pmb{W}}\right) + j\pmb{H}_{r,\ell}\operatorname{Im}\left(\Delta _{\pmb{W}}\right) \right]  \Bigg\}
	\end{aligned}
\end{equation}
leading to equation \eqref{eq:phi1_final}. 